\documentclass[12pt,twoside]{article}		
\usepackage{atmp}
\copyrightnotice{2001}{1}{165}{209}
\newcommand{\nc}{\newcommand}
\nc{\al}{\alpha}
\nc{\be}{\beta}
\nc{\de}{\delta}
\nc{\De}{\Delta}
\nc{\ga}{\gamma}
\nc{\ka}{\kappa}
\nc{\La}{\Lambda}
\nc{\la}{\lambda}
\nc{\Om}{\Omega}
\nc{\om}{\omega}
\nc{\si}{\sigma}
\nc{\Si}{\Sigma}
\nc{\ta}{\theta}
\nc{\va}{\varphi}
\nc{\ve}{\varepsilon}
\nc{\ze}{\zeta}
\nc{\ca}{{\mathcal A}}
\nc{\cb}{{\mathcal B}}
\nc{\cC}{{\mathcal C}}
\nc{\ce}{{\mathcal E}}
\nc{\ch}{{\mathcal H}}
\nc{\co}{{\mathcal O}}
\nc{\pa}{\partial}
\nc{\we}{\wedge}
\nc{\bC}{\mathbb C}
\nc{\bP}{\mathbb P}
\nc{\bR}{\mathbb R}
\nc{\bZ}{\mathbb Z}
\nc{\f}{\frac}
\nc{\sq}{\sqrt}
\nc{\iy}{\infty}
\nc{\op}{\oplus}
\nc{\ti}{\times}
\nc{\na}{\nabla}
\nc{\sub}{\subset}
\nc{\ra}{\rightarrow}
\nc{\ul}{\underline}
\nc{\lef}{\langle}
\nc{\ri}{\rangle}

\newtheorem{theorem}{Theorem}
\newtheorem{lemma}[theorem]{Lemma}
\newtheorem{proposition}[theorem]{Proposition}

\begin{document}

\title{Partitioning of electric and magnetic energy in SU(2) gauge theory}
\author{Clifford H. Taubes \footnote{Supported in part 
by the National Science Foundation}}

\address{Department of Mathematics\\
Harvard University\\
Cambridge, MA 02138}
\addressemail{chtaubes@math.harvard.edu}
\pagestyle{myheadings}
\markboth{PARTITIONING OF ELECTRIC AND MAGNETIC ENERGY...}{CLIFFORD HENRY TAUBES}

\url{xxxx}

\section*{Introduction}

It has been known for over twenty years that the Coulomb
solution in Maxwellian electro/magneto-statics for a point charge at
the origin in 3-space is unstable when embedded in a non-abelian gauge
theory with large coupling constant.  A version of this phenomenon was
first seen by Mandula \cite{Ma} and other manifestations of it were
described in some detail by Sikivie and Weiss \cite{SW1}-\cite{SW3}, Jackiw,
Jacobs and Rebbi \cite{JKR}, Jackiw and Rossi \cite{JR} and in Jackiw's
lecture notes, \cite{J}. \linebreak
\newpage

\pagestyle{myheadings}
\markboth{\hfil{\sc \small C.H. TAUBES\hspace*{20mm}}\hfil}%
{\hfil{\sc \small PARTITIONING of ELECTRIC and MAGNETIC ENERGY}\hfil}

\noindent
The instability arises more or less for the
following reason: The Maxwell\-ian Coulomb solution has no magnetic
field.  However, in a non-abelian gauge theory, static electric and
magnetic fields interact; and so it is energetically favorable at
large coupling to have a magnetic field that suppresses the coulomb
like behavior of the electric field.  This instability has nothing to
do with the singular nature of a point charge distribution; it occurs,
for example, when the charge is uniformly distributed over a ball.
This said, the Quixotic purpose of this article is to shed light on
the dependence of the minimal energy on the coupling constant for the
afore mentioned uniform charge distribution.  Of particular interest
are the relative contributions to this minimal energy from the
electric and magnetic fields.  Theorem 1, below, reports the results
on these questions.

Of further interest are the functional forms of the electric and
magnetic fields for an energy minimizer.  However, as their detailed
behavior resists analysis, the study here concentrates instead on the
behavior of these fields for an energy critical point that minimizes
energy under a restrictive hypothesis.  Even so, the energy of this
critical point is very close to the minimum energy and may well equal
the minimum energy.  In any event, the long range behavior of the
electric and magnetic potentials for this other critical point are
described below in some detail with Theorem 2 summarizing most of the
salient features.
	
Here is the background for the story: Let $su(2)$ denote the Lie algebra
of $SU(2)$, thus, the vector space of $2 \ti 2$ complex valued,
anti-hermitian and traceless matrices.  Now, let $A$ denote an $su(2)$
valued 1-form, and let $B_A \equiv *(dA + A \we A)$ denote its `magnetic
field'.  Here, $*$ is the Hodge star isomorphism from 2-forms to
1-forms; for example $*(dx^1\we dx^2)=dx^3$. Introduce a norm on $su(2)$ by
the requirement that its square to send $\tau\in su(2)$ 
to $|\tau|^2 \equiv -2\;$trace$\;(\tau^2)$; 
then use the latter with the Euclidean norm on $T^*\bR^3$ to
defined the norms of $su(2)$ valued 1-forms.

This done, the interest here is with the vector space, $\ca$, of $su(2)$
valued 1-forms A which vanish where $r \leq 1$, are smooth where $r > 1$,
have locally square integrable first derivatives and are such that
$|B_A|^2$ is integrable over $\bR^3$.  
Of special interest are the $su(2)$-valued
1-forms that minimize the energy functional, 
$\ce: \ca \ra (0, \iy)$, whose definition follows.

To define $\ce$, it is necessary to first digress with an introduction
to the covariant derivative, $\na_A$, defined by $A \in \ca$.  This
derivative sends an $su(2)$ valued function $\Psi$ to the $su(2)$
valued 1-form $\na_A\Psi = d\Psi + A \Psi -\Psi A$.  The covariant
derivative of $\Psi$ in the direction of a vector $v$ is obtained by
contracting $v$ with $\na_A\Psi$.  The covariant derivative defines
the covariant Laplacian, $\na_A$, which is the sum of the second
powers of the covariant derivatives in each of the 3 coordinate
directions.

Now, let $\tau^1 \in su(2)$ denote 
a fixed element with norm 1 and let $\rho$
denote the $su(2)$ valued function that equals $3 (4\pi)^{-1} \tau^1$ on the
unit ball in $\bR^3$ and vanishes on the complement of this ball.  Then
each $A\in\ca$ a determines a unique $su(2)$ valued function, 
$\Psi_A$, that obeys
$$
-\na_A \Psi_A = \rho \eqno(1) 
$$
and is such that both $|\na_A\Psi_A|$ and $|\Psi_A|^3$ square
integrable over $\bR^3$.  Standard calculus of variation techniques can be
used to argue both for the existence and uniqueness of $\Psi_A$ and to study
its dependence on $A$.

Given the preceding, specify a positive number $g$, the coupling
constant; and then define $\ce$ by the rule
$$
\ce(A) \equiv 2^{-1} g^{-2} \int |B_A|^2 
+ 2^{-1} g^2 \int |\na_A\Psi_A|^2 .  \eqno(2) 
$$
Here, and below the
notation is such that the integral sign with no indication of domain
or measure denotes integration over R3 with the Euclidean volume
element.
	
A critical point of $\ce$ is, by definition, 
an $su(2)$ valued 1-form $A$ with the property that
$$
\f{d}{dt} \ce(A + t a) |_{t=0} = 0\eqno(3)
$$
for all $a \in\ca$.  A critical point $A$ is called 
a local minimizer of $\ce$ if 
$$
\ce(A + t a) \geq \ce(A) \eqno(4) 
$$
for every $a$ and, given $a$, for all $t$ in some neighborhood 
of $0 \in\bR$.  A global minimizer of $\ce$ obeys (4) for all $a \in
\ca$ and $t \in\bR$.

With regards to the choice of ball radius 1 and charge normalization 
$\int|\rho| = 1$ in the definitions 
of $\ca$ and $\ce$, note that changes of either can
be absorbed by a combination of rescaling $\ce$ and $g$.  Indeed, if the
corresponding variational problem is posed in the ball of radius $R$,
with charge density $\rho$ such that 
$\int|\rho| = q$, and with coupling constant $g$, 
then the corresponding energy infimum is equal to the product of
$q^{1/2} R^{-1}$ times the energy infimum 
for the $R = 1$, $q = 1$ and coupling
constant $g q^{1/2}$ version of the functional in (2).  In any event, this
rescaling property justifies the focus here on the use of the unit
ball and $\int|\rho| = 1$ charge in the definitions of $\ca$ and $\ce$.
	
By the way, the 1-forms in $\ca$ are required to vanish inside the unit
ball for three reasons of which pay small homage to the underlying
physics.  First, high energy particle physics experiments observe that
the non-abelian magnetic fields associated to the subnuclear forces
are small at short distances.  Second, physics would, in any event,
give the charge density $\rho$ as a functional of some other field (say a
spinor with values in $\bC^2$), and the latter would, perforce, obey an
equation that also involved the 1-form $A$.  Because such an auxiliary
equation is not considered here, a formulation that allows $A$ to vary
in the ball lacks a certain logical consistency.  Finally, allow $A$ to
vary where $\rho\neq 0$ and there is no invariant notion of charge density
to allow energy comparisons.  The latter point underlies statements in
some of the afore-mentioned references to the effect that the Coulomb
solution is unstable for all values of the coupling constant.

Note that this variational setting enjoys a restrictive sort of gauge
invariance.  Indeed, if $A$ is in $\ca$ and $h: \bR^3 \ra SU(2)$ is
the identity matrix on the unit ball, then the `gauge equivalent'
1-form $h^{-1}Ah + h^{-1}dh$ is also in $\ca$ and $\ce(A) = e(h^{-1}Ah
+ h^{-1}dh)$.

As a parenthetical remark, note that the variational setting here can
be formulated in a completely gauge invariant manner.  To do so, first
change the definition of $\ca$ so that the requirement that $A$ vanish
in the ball is replaced by the requirement that $B_A$ vanish in the
ball.  This done, make the following requirements on $\rho$: First, it
should vanish outside the ball and its norm should equal $3/(4\pi)$
inside.  Second, require that $\na_A\rho = 0$ where $\rho\neq 0$.
This done, then the energy $\ce$ can be considered to be a function of
pairs $(A, \rho)$, subject to the preceding constraints.  As such,
this energy is gauge invariant since $(h^{-1}Ah + h^{-1}dh, h^{-1}\rho
h)$ and $(A,\rho)$ have the same energy for all smooth $h: \bR^3 \ra
SU(2)$.
	
The author knows the explicit form of only one critical point of $\ce$ (up
to gauge equivalence), this being the gauge transforms of the
`coulomb' solution which has $A = 0$ with the corresponding $\Psi_A$
given in terms of the radial coordinate $r$ on $\bR^3$ as

\begin{itemize}
\item
$\Psi_{A=0} =  (8\pi)^{-1} (3 - r2) \tau^1$ where  $r \leq 1$. \hfill (5)
\item
$\Psi_{A=0} = (4\pi)^{-1} r^{-1} \tau^1$ where $r \geq 1$.
\end{itemize}
The energy, $\ce(0)$, of the Coulomb solution is equal 
to $3 g^2 (40\pi)^{-1}$.  
	
With the preceding understood, consider:

\begin{theorem}		
For each $g > 0$ the corresponding version of the functional $\ce$ achieves
its global   minimum on $\ca$.  
When $g \leq (\sq{2} \pi)^{1/2}$ this minimum is attained
only on the Coulomb solution and its  
gauge transforms.  However, when  $g > (6\pi)^{1/2}$ then the Coulomb
solution is no longer the  
minimizer of $\ce$.  Moreover, there exists a g-independent constant $c$
such that when $g > (6\pi)^{1/2}$, then
$$
g^2 (40\pi)^{-1} + c^{-1} g \leq \inf_{\ca} \ce 
\leq g^2 (40\pi)^{-1} + c g.\eqno(6)
$$
In addition, if, for such $g$, the corresponding version of $\ce$ achieves
its infimum at $A \in\ca$, then

\begin{itemize}
\item 
$g^2 (20\pi)^{-1} \leq \int_{r\leq 1} |\na_A\Psi_A|^2$. \hfill $(7)$
\item
$c^{-1} g  \leq g^2 \int_{r\geq 1} |\na_A\Psi_A|2 \leq c g$.
\item
$c^{-1} g  \leq g^{-2} \int_{1\leq r\leq 1+c/g} |B_A|^2$   
and   $g^{-2} \int_{1\leq r} |B_A|^2 \leq c g$. 
\item
$|\Psi_A|<|\Psi_{A=0}|$ everywhere, and  
$|\Psi_A| \leq c g^{-1/2}  r^{-1}$ where  $r\geq 2$.
\end{itemize}
\end{theorem}

Remark that the factor $g^2 (40\pi)^{-1}$ appears in (6) and in the first
point of (7) because $A = 0$ in the unit ball and $(40\pi)^{-1}$ is the
minimum over the set of solutions in the unit ball to the equation -
$\De_{A=0}\Psi = \rho$ of the functional that sends 
$\Psi$ to $\int_{r\leq 1} |\na_{A=0}\Psi|^2$.  By the
way, note in the second point of (7) that a substantial fraction of
the magnetic energy is contributed from the shell of thickness $\co(g^{-1})$
that surrounds the unit ball.  Note also that the results in Theorem 1
do not change in any substantial way if the charge distribution $\rho$ in
(1) is allowed to vary in the ball as 
$\rho_0 \tau^1$ with $\rho_0$ positive with integral 1.
	
As remarked above, the author has little to say about the detailed
point to point behavior of the fields $(A, \Psi_A)$ when $A$ is an
absolute minimizer of a large $g$ version of $\ce$.  On the otherhand,
quite a bit can be said about these fields for the absolute minimizer
of $\ce$'s restriction to a certain subset, $\ca_0 \sub \ca$.  Here,
$A \in \ca_0$ when $A$ is gauge equivalent to $A_0 = \al \sin \ta d\va
\tau^2$ where $\al$ is a function on $\bR^3$, $\ta$ and $\va$ are the
standard spherical coordinates and $\tau^2 \in su(2)$ has unit length
and is orthogonal to $\tau^1$.  In this regard, the critical points of
$\ce$'s restriction to $\ca_0$ are also unrestricted critical points
of $\ce$.  This understood, the next theorem describes the minimizers
of $\ce$'s restriction to $\ca_0$.

\begin{theorem}		
For each $g > 0$, there is a function $\al$ on $\bR^3$,
unique up to multiplication by $\pm 1$, and charcterized by the fact
that the global minimizers of $\ce$'s restriction to $\ca_0$ are all gauge
equivalent to $A \equiv\al \sin \ta d\va\tau^2$.  
In this regard, $\al = 0$ when $g \leq (6\pi)^{1/2}$, 
but not so when $g > (6\pi)^{1/2}$.  In addition, there is a
constant, $c \geq 1$, with the following significance: For $g >
(6\pi)^{1/2}$, the energy $\ce(A)$ obeys $(6)$ and $A$ 
with its corresponding $\Psi_A$ obeys $(7)$.  Moreover,
\begin{itemize}
\item
$\Psi_A = \Psi\tau^1$ with $\Psi$ a positive function on $\bR^3$. 
\item
$\al = f \sin \ta$ with $|f| > 0$ where $r > 1$.  
\item
At points where $r \geq c$,
\begin{itemize}
\item[{\rm a)}]	
$\Psi \leq \sq{2} g^{-2}e^{-g^{1/4}/c} r^{-1}$.
\item[{\rm b)}]  
$\al = f r^{-1} \sin \ta$ with  $|f| \geq g^{1/2}/c$.
\end{itemize}
\item
At points where  $r \gg c$,
\begin{itemize}
\item[{\rm a)}]  
$\Psi = \sq{2} g^{-2} (e_0 + m_{\Psi}) r^{-1}$  
where $e_0$ is a positive constant less than $e^{-g^{1/4}/c}$
and $|m_{\Psi}| \leq c r^{-1}$. 
\item[{\rm b)}]  
$\al = (c_0 + m_{\al}) r^{-[(9-8e_0^2)^{1/2}-1]} \sin\ta$ 
where $c_0$ is a positive constant and  $|m_{\al}| \leq c r^{-1}$.  
\end{itemize}
\end{itemize}
\end{theorem}

Of particular interest to the author with regards to Theorem 2 is the
fact that the electric potential, $\Psi$, although Coulomb like at large
distance, has an effective charge $\sq{-2} g^{-2} e_0$ that is absolutely
microscopic at large $g$.  Meanwhile, the magnetic potential, $\al$, falls
to zero rather like the field of a dipole charge with dipole moment of
size $\co(g^{1/2})$.

As with the statement of Theorem 1, the conclusions of Theorem 2 are
not changed substantially if the charge distribution $\rho$ in (1) is
allowed to vary in the ball as $\rho_0 \tau^1$ with $\rho_0$ positive
with integral 1.  This said, only the $\rho_0 = 3 (4\pi)^{-1}$ case is
discussed below.

The remainder of this article is occupied with the proofs of
Theorems 1 and 2.  In this regard, the assertions of Theorem 2 are
proved first as they are used in part to prove Theorem 1.  In
particular, the assertions of Theorem 2 are proved as various
propositions and lemmas in Sections a-j below.  These are all
summarized in Section k to tie up the argument for Theorem 2.  Section
k then ends with the proof of Theorem 1.

Before starting, note that the author benefitted at an early
stage of this project from conversations with R. Scott.

\section*{a)  A reformulation of the variational problem on $\ca_0$}

	The variational problem given by $\ce$'s restriction to the
set $\ca_0$ of $su(2)$-valued 1-forms gauge equivalent to $\al \sin
\ta d\va\tau^2$ with $\al$ a function on $\bR^3$ can be reformulated
as follows:
	
Let $g \geq 0$ be a constant, and, for now, let $\rho$ be a function with
compact support where the distance, $r$, to the origin is less than 1
and with integral equal to 1.  Now consider an equation for a $C^0$
function $\al$ of the first two 
of the spherical coordinates $(r, \ta)$ on $\bR^3$
which is given as follows: First, introduce the unique, $C^1$ solution 
$\psi\equiv\psi[\al]$ on $\bR^3$ to the equations

\begin{itemize}
\item
$-\De\psi + r^{-2} \al^2 \psi = \rho$  where $r > 1$. \hfill (8)
\item
$\lim_{r\ra\iy} = 0$.
\end{itemize}
Here, $\psi$ is a function of $(r, \ta)$ too, and 
$$\De = r^{-2} (r^2 (\,\cdot\,)_r)_r + (\sin\ta)^{-1} 
(\sin \ta (\,\cdot\,)_{\ta})_{\ta}$$ 
is the standard Laplacian.  Then, with 
$\psi$ understood, require that

\begin{itemize}
\item
$\al = 0$ where $r \leq 1$. \hfill (9)
\item
$- \al_{rr}- r^{-2} ((\sin \ta)^{-1} (\sin \ta\al)_{\ta})_{\ta} 
- g^4 \psi^2 \al = 0$ where $r > 1$.
\item
$r^{-2} \al$ is square integrable on $\bR^3$.
\end{itemize}
Here, and also below when ambiguities are unlikely, the partial
derivative of a function, $f$, with respect to $r$ is written as $f_r$ and
with respect to $\ta$ as $f_{\ta}$.  When ambiguity can arise, these
derivatives are written below as $\pa_rf$ and $\pa_{\ta}f$, respectively.

Of special interest are those functions $\al$ for that are absolute
minimizers of the functional 
\begin{multline*}
\ce_0(\al)\equiv 2^{-1} g^{-2}\int r^{-2}(\al_r^2 
+ r^{-2} (\sin \ta)^{-2} (\sin \ta\al)_{\ta}^2) \\
+  2^{-1} g^2 \int (|\na\psi|^2 + r^{-2} \al^2 \psi^2) .\tag{10}
\end{multline*}
Note that $\ce_0(\al) = \ce(\al \sin \ta d\va\tau^2)$ 
with $\ce$ as in (2), so $\ce_0$ is the restriction 
of $\ce$ to those $A \in \ca$ of the form $\al \sin \ta d\va\tau^2$.
	
Note that the introduction of the 1-form 
$a \equiv \al\sin \ta d\va$ allows the $r\geq 1$ 
parts of (8) and (9) to be rewritten as	

\begin{itemize}
\item
$-\De\psi + |a|^2 \psi = 0$, \hfill (11)
\item  	
$-\De a - g^4 |\psi|^2 a = 0$.
\end{itemize}

In terms of the 1-form $a$, the energy $\ce_0$ becomes
$$
\ce_0 = 2^{-1} g^{-2} \int |\na a|^2  
+ 2^{-1} g^2 \int (|\na\psi|^2 + |a|^2 |\psi|^2) .\eqno(12)
$$
	
Note that $\ce_0$ in (10) defines a bonafide functional on the space,
$\cC_0$, of continuous functions on $\bR^3$ that vanish on the unit
ball, are invariant under rotations about the axis defined by the
spherical angle $\va$ and are such that $r^{-2} (\al_r^2 + r^{-2}
(\sin \ta)^{-2} (\sin \ta\al)_{\ta}^2)$ has finite integral.
Moreover, techniques from the calculus of variations establish that
$\ce_0$ achieves its infimimum on $\cC_0$.  In addition, standard
elliptic regularity techniques (as found, for example, in Chapter 6 of
\cite{Mo}) justify the assertion that any critical point of $\ce_0$ on
$cC_0$ is smooth where $r > 1$.

\section*{b)  Some simple inequalities}

	The subsequent analysis exploits some basic observations that
concern $\ce_0$ and the solutions to (8) and (9).  In this regard, the
first observation is obtained by multiplying both sides of the top
equation in (11) by $\psi$ and then integrating the result over
$\bR^3$.  A subsequent integration by parts then finds that
$$
\int (|\na\psi|^2 + |a|^2 |\psi|^2) = \int\psi\rho. 
\eqno(13)
$$  
The second observation follows by contracting both sides of the bottom
equation in (11) with the 1-form a and then integrating the result
over $\bR^3$.  This done, an integration by parts finds
$$
g^{-2} \int |\na a|^2 = g^2 \int |a|^2 |\psi|2 .
\eqno(14)
$$
The third observation concerns a pair $\psi$ and $\psi'$ where these
functions solve the versions of the second that are defined by a
corresponding pair, $a$ and $a'$.  This understood, note the following:
\begin{multline*}
\text{{\it If  $|a| \geq |a'|$ everywhere with the inequality 
strict somewhere,}}\\ 
\text{{\it then $\psi <\psi' $ everywhere.}}
\tag{15}
\end{multline*}
Indeed, this follows from the maximum principle because the top
equation in (8) implies that 
$$
-\De(\psi- \psi') + (|a^2| - |a'|^2) \psi + |a'|^2 (\psi - \psi') = 0.
\eqno(16)
$$
In particular, this equation precludes a non-negative maximum for
$\psi-\psi'$.   

The observation in (15) implies that for any $a$, the corresponding
$\psi$ is sandwiched as 
$$
\psi_D < \psi\leq\psi_{\rm Coul} ,\eqno(17)
$$
where $\psi_{\rm Coul}$ is the solution to 
$- \De\si = \rho$ on $\bR^3$ which decays to zero as
$r \ra\iy$, while $\psi_D$ is the  solution to $- \De\si = \rho$
in the ball where $r \leq 1$ which vanishes at $r =1$.  
For example, in the case where  
$\rho = \rho_0 = 3 (4\pi)^{-1}$ where $r \leq 1$, 

\begin{itemize}
\item
$(8\pi)^{-1} (1 - r^2) < \psi\leq (8\pi)^{-1} (3 - r^2)$ \quad
where $r \leq 1$. \hfill (18)
\item
$0 < \psi\leq (4\pi)^{-1} r^{-1}$ \hspace*{35mm} where $r \geq 1$.
\end{itemize}

By the way, note that (13) and (18) imply that 
$$
(40 \pi)^{-1} < 2^{-1} \int (|\na\psi|^2 + |a|^2 |\psi|^2) 
\leq 3 (20 \pi)^{-1} \eqno(19)
$$
in the case where $\rho$ is the constant $3 (4\pi)^{-1}$.

\section*{c)  The coulomb solution}

The Coulomb solution to (8) and (9) has $\al = 0$ and 
$\psi\equiv\psi_0$ given by
$$
\psi_{\rm Coul}(x) = (4\pi)^{-1} \int |x - (\cdot)|^{-1} \rho.
\eqno(20)
$$
For example, when $\rho$ has the constant value $3 (4\pi)^{-1}$ in the unit
ball and is zero outside, then $\psi_0$ is  
equal to $(8\pi)^{-1}(3 - r^2)$ in the ball and $(4\pi)^{-1}r^{-1}$ outside.
As the following lemma attests, the  Coulomb solution, 
$\al\equiv 0$, is the minimizer of $\ce_0$ when $g$ is small and
not the minimizer when $g$ is  large.

\begin{lemma}			
When $g \leq g_0,$ then $\al = 0$ is the absolute minimizer of $\ce_0$
on $\;\cC_0,$ and when $g > g_0,$ it is not.  In the case where $\rho =
3 (4\pi)^{-1}$ in\-side the ball and zero outside, $g_0 = (6\pi)^{1/2}$.
\end{lemma}

The proof of this lemma exploits two fundamental inequalities that are
also used elsewhere  in the paper:

\hfill (21)
{\it
\begin{itemize}
\item
Suppose that $f$ is a function of $r$ that vanishes where $r \leq 1$
and is such that $f_r$ is square integrable on $[1, \iy)$ with respect 
to $dr$.  Then $\int_{r\geq 1} r^{-2} f^2 dr\leq 4 \int_{r\geq 1} f_r^2 dr$. 
\item
Suppose that $f$ is a function of $r$ that vanishes in the limit as 
$r\ra\iy$ and is such that $rf_r$ is square integrable with respect 
to $dr$.  Then, $\int_{r\geq 1} f^2 dr \leq 4 \int_{r\geq 1} f_r^2 r^2 dr$.
\item
Suppose that $f$ is a function of the spherical angle $\ta$ 
and is such that  $(\sin \ta)^{-1} (\sin \ta f)_{\ta}^2$ is integrable 
on $[0, \pi]$.  Then  $\int_{0\leq\ta\leq\pi} \sin \ta f^2 d\ta$ $\leq 
2^{-1}\int_{0\leq\ta\leq\pi} (\sin \ta)^{-1} (\sin \ta f)_{\ta}^2 d\ta$. 
\end{itemize}
} 
To argue for the first inequality, write $r^{-2} dr$ as $-d(r^{-1})$
on the left hand side of the integral, integrate by parts and then use
the triangle inequality.  The second inequality follows by directly
integrating by parts on the left hand side of the integral and then
employing the triangle inequality.  The third inequality follows from
the fact that the corresponding Sturm-Liouville operator has smallest
eigenvalue 2.  Note for future reference that $f = \sin \ta$ is the
corresponding eigenfunction.	

\medskip\noindent
{\it Proof of Lemma} 3. \  
Consider first the argument that $\al = 0$ is the
minimizer of $\ce_0$ when $g$ is small.   
For this purpose, appeal to the first point in (21) and also (13) to
derive the inequality  
$$
\ce_0(\al) \geq (9/8) g^{-2} \int r^{-4} \al^2 
+ 2^{-1} g^2 \int_{r\leq 1}\rho\psi.\eqno(22)
$$
Now, multiply the top line of (8) by the Coulomb solution 
$\psi_{\rm Coul}$ and integrate the result over $\bR^3$ to 
find that 
$$
\int_{r\leq 1}\rho\psi + \int r^{-2} \al^2 \psi\psi_{\rm Coul} 
= \int_{r\leq 1}\rho\psi_{\rm Coul} .\eqno(23)
$$
Then, since $\ce(0) = 2^{-1}\int_{r\leq 1}\rho\psi_{\rm Coul}$, this
last equality, (17) and (22) imply that 
$$
\ce_0(\al) - \ce_0(0) \geq (9/8) g^{-2} \int r^{-4} \al^2 
- 2^{-1} g^2 \int r^{-2} \al^2 \psi_{\rm Coul}^2 .\eqno(23)
$$
Finally, $\rho\geq 0$, so the maximum principle finds $\psi_{\rm
Coul}> 0$ everywhere and $\psi_{\rm Coul} \geq m r^{-1}$ where $r\geq 1$; 
here $m = \min_{r=1} \psi_{\rm Coul}$.  This understood, then (23)
asserts that the Coulomb solution is the minimizer when $g \leq
3^{1/2} (2m)^{-1/2}$.  For example, when $\rho = 3 (4\pi)^{-1}$, the
Coulomb solution is the minimizer when $g \leq (6\pi)^{1/2}$.

To see that $\al = 0$ is not the minimizer of $\ce_0$ when $g$ is
large, it is necessary only to prove that the Hessian of the large g
versions of $\ce_0$ at $\al = 0$ is not a positive semi-definite
quadratic form on $\cC_0$.  In this regard, note that this Hessian
assigns to each $\be\in\cC_0$ the number
$$
\ch_0(\be) \equiv g^{-2} \int_{r\geq 1} r^{-2} 
(\be_r^2 + r^{-2} (\sin \ta)^{-2} (\sin\ta\be)_{\ta}^2) 
- g^2 \int_{r\geq 1} r^{-2} \be^2 \psi_{\rm _Coul}^2 .\eqno(24)
$$
Now, according to the maximum principle, $\psi_{\rm _Coul} \leq M
r^{-1}$ where $M$ denotes the maximum value of  
$\psi_{\rm Coul}$ on the $r = 1$ sphere.  This understood, then
$$
\ch_0(\be) \leq g^{-2} \int_{r\geq 1} r^{-2} \left(\be_r^2 + r^{-2} 
(\sin \ta)^{-2} (\sin \ta\be)_{\ta}^2\right) 
- g^2 M^2 \int_{r\geq 1} r^{-4} \be^2.\eqno(25)
$$
	
With (25) understood, fix some small $\ve > 0$ and take $\be$ to be
zero where $r \leq 1$ and to equal the function $(r^{(1-\ve)/2} - 1)
\sin \ta$ where $r \leq 1$.  This done, a calculation finds that
$$
\ch_0(\be) < (8\pi/3) \ve^{-1} (g^{-2} 9/8 - g^2 M^2 + \co(\ve)),
\eqno(26)
$$
which is negative for small $\ve$ provided that $g \geq 3^{1/2}
(2M)^{-1/2}$.

\section*{d)  Some energy inequalities}

	The next proposition gives a first indication of the large $g$
behavior of various parts of the infimum of $\ce_0$.  In the statement
of this proposition and in the discussions of the subsequent sections
of this paper, the function $\rho$ in (8) is implicitly that which
vanishes outside the unit ball and equals the constant $3 (4\pi)^{-1}$
inside.

\begin{proposition}			
There is a constant $c_1$ with the following significance:  Given $g >
(6\pi)^{1/2},$ suppose  
that $\al$ is a minimizer of $\ce_0$.  Then $\al$ 
and its corresponding $\psi$ obey  
\begin{itemize}
\item
$(20\pi)^{-1} < \int_{r\leq 1} |\na\psi|^2 < (20\pi)^{-1} + c_1 g^{-1}$.
\item
$\int_{r>1} |\na\psi|^2 < c_1 g^{-1}$ .
\item
$\int|a|^2 \psi^2 <c_1 g^{-1}$.
\item
$\psi < c_1 g^{-1/2} r^{-1}$  where $r > 2$.
\end{itemize}
\end{proposition}

The remainder of this section is occupied with the

\medskip\noindent
{\it Proof of Proposition}\; 4. \  
The proof starts with the following observa\-tion:  
Let $\si$ be a function on 
the unit ball that obeys $-\De\si = 3 (4\pi)^{-1}$.  Then, 
$$
\int_{r\leq 1} |\na\si|^2 \geq (20\pi)^{-1} , \eqno(27) 
$$
and is an equality if and only if $\si = (8\pi)^{-1} (m - r^2)$ with
$m \in \bR$.  Indeed, this follows by writing $\si$ as a sum of
products of functions of $r$ times spherical harmonics.  The preceding
inequality implies the left hand inequality in the first point of the
proposition.

There are two parts to the proofs of the remaining assertions.  The
first part below proves the right hand inequality in the first point,
and both the second and third points of Proposition 4.  The second
part proves the final point.

\medskip\noindent
{\it Part} 1. \ 
Let $E(g)$ denote the infimum of the $g$-version of $\ce_0$.
What with (14), a bound by $g^2 (40\pi)^{-1} + c_1g$ on $E(g)$ gives the right
hand inequality in the first point of Proposition 4 plus the next two
points in the proposition.  This step establish such an upper bound
for $E(g)$.

For this purpose, fix, $\ve > 0$ and a non-decreasing function $\be$ on $[0,
\iy)$ that has value 0 on $[0, 1]$, equals 1 on $[1 + \ve, \iy)$ and obeys
$\be'< 2/\ve$.  This done, fix $\la > 1$ and fix a smooth function $\chi: [0,
\pi] \ra [0, 1]$ that obeys $\chi(\ta) = \chi(\pi - \ta)$, 
equals 1 on $[\la^{-1},\pi/2]$, 
equals $\la\ta$ for $\ta < (2\la)^{-1}$, and obeys $|\chi'| \leq 4 \la$.

Now set $a = \la\be(r) \chi(\ta) \sin \ta d\va$.  Let
$\psi\equiv\psi[\la,\ve]$ denote the corresponding solution to the
$a$-version of the first equation in (8).  Then, a straight forward
calculation finds that
$$
 E(g) \leq c g^{-2} \la^2 (\ve^{-1} + \ln(\la)) 
+ 3/(8\pi) g^2 \int_{r\leq 1} \psi  ;\eqno(28)
$$
here $c$ is a constant which is independent of $g$, $\ve$ and $\la$.

Hold onto (28) for the moment, and introduce the piece-wise continuous
1-form $a'$ that is equal $\la\chi \sin \ta d\va$ 
where $r \geq 1 + \ve$ and equal
to 0 where $r \leq 1 + \ve$.  Let $\psi'\equiv\psi'[\la, \ve]$ denote the
corresponding solution to the $a'$ version of (8).  By virtue of
(15), $\psi < \psi'$ since $|a| \geq |a'|$. Thus, (28) implies that
$$
E(g) \leq c g^{-2} \la^2 (\ve^{-1} + \ln(\la)) 
+ 3/(8\pi) g^2 \int_{r\leq 1} \psi' .\eqno(29)
$$
	
To make further progress, introduce the function $\ul{\rho}$ which is
defined to equal $3/(4\pi)$ where $r \leq 1 + \ve$ and to equal 0
where $r > 1 + \ve$.  Let $\ul{\psi}$ denote the corresponding
solution to the equation
$$
-\De\ul{\psi}+ |a'|^2 \ul{\psi} = \ul{\rho}.
\eqno(30)
$$
on $\bR^3$ which decays to zero as $r \ra\iy$.  As $\psi'$ solves the
analogous equation with $\rho'$ replacing $\ul{\rho}$ and since
$\ul{\rho}\geq \rho'$, one has
$$
-\De(\psi' - \ul{\psi}) + |a'|^2 (\psi' - \ul{\psi}) \leq 0 
\eqno(31)
$$
on the whole of $\bR^3$.  In particular, (31) with the maximum
principle implies that $\psi' <\ul{\psi} $.  Thus,  
$$
 E(g) \leq c g^{-2} \la^2 (\ve^{-1} + \ln(\la)) 
+ 3/(8\pi) g^2 \int_{r\leq 1} \ul{\psi}.  \eqno(32)
$$
Moreover, since the integral of $\ul{\psi}$ over the radius 1 ball is
less than its integral over the ball of radius $1 + \ve$, the
preceding inequality immediately give
$$
E(g) \leq c g^{-2} \la^2 (\ve^{-1} + \ln(\la)) 
+ 3/(8\pi) g^2 \int_{r\leq 1+\ve} \ul{\psi}.\eqno(33)
$$
	
Now, the next step is to rescale the expression on the far right in
(33) so that the integral is over the radius 1 ball again.  For this
purpose, introduce the function $\eta$ which assigns to the point $x\in\bR^3$
the value $\eta(x) \equiv (1 + \ve)^{-2} \ul{\psi}((1 + \ve) x)$.  
In terms of $\eta$, the inequality in (33) reads
$$
E(g) \leq c g^{-2} \la^2 (\ve^{-1} + \ln(\la)) 
+ 3/(8\pi)(1+\ve)^5 g^2 \int_{r\leq 1} \eta.
\eqno(34)
$$
Moreover, the function $\eta$ obeys the equation
$$
-\De\eta + |b|^2 \eta = \rho,\eqno(35)
$$
for the case with b given by $\la\chi\sin \ta d\va$ 
where $r \geq 1$ and $b = 0$ where $r \leq 1$. 
	
To complete the proof of the second point of Proposition 4, it is
necessary to bound the size of $3/(8\pi) \int_{r\leq 1} \eta$.  
For this purpose,
note that this integral is equal to the supremum over all $C^1$ functions
u which decay to zero as $r \ra\iy$ of the functional
$$
e(u) \equiv 3/(4\pi)\int_{r\leq 1}u-2^{-1}\int(|\na u|^2+|b|^2u^2).
\eqno(36)
$$
In this regard, there is a constant $c_0 > 0$ which 
is independent of $\la$ and is such that when $\la\geq 1/100$, 
the following is true:   Let $u$ be any $C^1$ 
function of $\ta\in [0, \pi]$.  Then 
$$
\int_{[0,\pi]} (u_{\ta}^2 + \la^2 \chi^2 u^2) 
\sin \ta d\ta \geq c_0 \la^2 \int_{[0,\pi]} u^2 \sin \ta d\ta. 
\eqno(37)
$$
Thus, for any $C^1$ function u as in (36), one has
$$
e(u) \leq 3/(4\pi) \int_{r\leq 1} u 
- 2^{-1} \int|\pa_ru|^2 - 2^{-1} c_0 \la^2 \int_{r\geq 1} r^{-2} u^2 .
\eqno(38)
$$
And, (38) implies that
\begin{align*}
& 3/(8\pi) \int_{r\leq 1} \eta\tag{39}\\
& \leq \sup_u \left\{3/(4\pi) \int_{r\leq1} u - 2^{-1} \int|\pa_ru|^2 
- 2^{-1} c_0 \la^2 \int_{r\geq 1} r^{-2} u^2\right\} .
\end{align*}
	
The point now is that the supremum on the right hand side of (39) can
be calculated explicitly because a maximizing function, $u_0$, can be
exhibited: For this purpose, introduce the number $p = 2^{-1}(1 + (1 + 4
c_0 \la^2)^{1/2})$, and then
\begin{itemize}
\item
$u_0 = (8\pi)^{-1} (2/p + 1 - r^2)$ \qquad where $r \leq 1$. \hfill (40)
\item
$u_0 = (4\pi)^{-1} p^{-1} r^{-p}$ \hspace*{20.5mm} where $r \geq 1$.
\end{itemize}
Given u0, the value of the supremum on the right hand side of (40) can
be readily computed to be $(40\pi)^{-1} (1 + 5/p)$.  
Since $1/p \leq c_0^{1/2} \la^{-1}$, this means that 
$$
3/(8\pi) \int_{r\leq 1} \eta
\leq (40\pi)^{-1} (1 + 5 c_0^{-1/2} \la^{-1}),
\eqno(41)
$$
and
$$
E(g) \leq c g^{-2} \la^2 (\ve^{-1} + \ln(\la))
+ (40\pi)^{-1} g^2 (1 + \ve)^5 (1 + 5 c_0^{-1/2} \la^{-1}).
 \eqno(42) 
$$	
This last inequality holds for any choice of $\ve > 0$ and $\la > 1$,
and so their values will be  
chosen to make the left hand side of (42) small.  For this purpose,
the first observation is that when  
$\ve < 1/2$ and $\la > 5/c_0^{1/2}$, 
then the expression on the right hand side of
(42) is no smaller than  
$$
 E(g) \leq c g^{-2} \la^2 (\ve^{-1} + \ln(\la)) 
+ (40\pi)^{-1} g^2 (1 + 5 c_0^{-1/2} \la^{-1}) + g^2 \ve .\eqno(43)
$$
Moreover, the left hand side of (43) is no greater than its value at 
$\ve= \surd c g^{-2} \la$; thus 
$$
E(g) \leq 2\surd 2 c \la + c g^{-2} \la^2 \ln(\la) 
+ (8\pi)^{-1} g^2 c_0^{-1/2} \la^{-1} + (40\pi)^{-1} g^2 .\eqno(44)
$$
	
Finally, $E(g)$ is no greater than the value of the left hand side of
(44) in the case $\la = g$ which  gives the bound
$$
E(g) \leq (40\pi)^{-1} g^2 + c' g.\eqno(45)
$$
Here, $c'$ is a constant which is independent of $g$.

\medskip\noindent
{\it Part} 2. \   
The assertion in the fourth point of Proposition 4 follows using the maximum 
principle with the equation in the second point of (8) given that
there is a $g$-independent constant $c'_1$ such that
$$
|\psi|(x) \leq c'_1 g^{-1/2}\qquad\text{where }\;r \geq 2,\eqno(46)
$$
To prove this last inequality, consider that (46) implies the equality
$$
\psi(x)^2 = (2\pi)^{-1} \int |x - (\cdot)|^{-1} 
(\rho\psi - |\na\psi|^2 - |a|^2 \psi^2) .\eqno(47)
$$
Store this last equation momentarily to fix a smooth function $\be: [0,
\iy) \ra [0, 1]$ which equals 1 on  $[0, 1]$ and vanishes on 
$[3/2, \iy)$.  Promote $\be$ to a function, $\ul{\be}$, on $\bR^3$
by setting $\ul{\be}(y) = \be(|y|)$.     
	
With $\ul{\be}$ understood, remark that (47) implies that
$$
\psi(x)2 \leq (2\pi)^{-1} \int |x - (\cdot)|^{-1} 
\ul{\be} (\rho\psi - |\na\psi|^2 - |a|^2 \psi^2) \eqno(48)
$$
since $\rho = 0$ where $\ul{\be}\neq 1$.  Then, an integration by parts finds
$$
\psi^2(x) \leq (4\pi)^{-1} \int |\De(\ul{\be} |x - (\cdot)|^{-1})| \psi^2.
\eqno(49)
$$
Now, if $|x| \geq 2$, then the right hand side of (49) is no greater than
$$
 c' \int_{1\leq r\leq 2} \psi^2,\eqno(50)
$$
where $c'$ is a constant which depends only on the particular choice
for $\be$.  Meanwhile, the integral in  (50) is bounded by 
$16 \int_{r\geq 1} |\na\psi|^2$ by virtue of the second point in
(21).  Thus, (49) and (50) imply  that
$$
\psi(x)^2 \leq c'' \int_{r\geq 1} |\na\psi|^2 \eqno(51)
$$
at all points $x$ with $|x| \geq 2$.  This last inequality and the
inequality in the second point to Proposition  4 completes 
the argument for the fourth point of Proposition 4.

\section*{e)  Some preliminary conclusions about $\al$ at large $r$}

The purpose of this subsection is to begin the study the pointwise
behavior of a minimizer, $\al$, of $\ce_0$.  The particular
observation in this section is summarized by

\begin{proposition}		
If $\al$ is a solution to $a$ $g > 0$ version of $(9),$ 
then $|\al|$ is uniformly bounded on $\bR^3$.   
Moreover, given $\ve > 0,$ there exists $r_{\ve}$ 
such that $\al <\ve $ when $r > r_{\ve}$.
Said differently, $a = \al \sin \ta d\va $ 
obeys $|a| \leq \ve r^{-1}$ when $r > r_{\ve}$.
\end{proposition}

\noindent
{\it Proof of Proposition} 5. \  
The derivation of the asserted bound requires a four step argument. 
 
\medskip\noindent	
{\it Step} 1. \   
Fix $r \geq 1$ and integrate both sides of the equation in the first
point of either (8) or (11) with respect to the standard spherical 
measure $d\Om\equiv \sin\ta d\ta d\va$ on the unit sphere.  After an  
integration by parts and multiplication by $r^2$, 
the resulting equation reads:
$$
- \pa_r r^2 \pa_r \int_{S^2} \psi(r, \cdot) d\Om 
+ \int_{S^2} r^2 (|a|^2 \psi)(r, \cdot) d\Om = 0.\eqno(52)
$$
Apply the maximum principle to this last equation to conclude that 
$$
\pa_r \int\psi(r, \cdot) d\Om < 0 \eqno(53)
$$
where $r \geq 1$.  

\medskip\noindent	
{\it Step} 2. \   
Now, integrate both sides of (52) with respect 
to the measure dr over the interval $[1, R]$ 
for any chosen $R \geq 1$.  The result is
$$
-\left(r^2 \pa_r \int\psi(r, \cdot) d\Om\right)_{r=R}  
+  \int_{1\leq r\leq R} |a|^2 \psi = -\int \psi_r(1, \cdot) d\Om= 1.
\eqno(54)
$$
Here, the right hand equality follows by integrating both sides of the
equation in the first point of (8) over the radius 1 ball.  
	
This last equation implies that
$$
\int |a|^2 \psi \leq 1 .\eqno(55)
$$

\medskip\noindent {\it Step} 3. \ 
Equation (55) implies that $|x -(\cdot) |^{-1} |a| \psi^2$ 
is integrable for any choice of $x \in
\bR^3$ with $|x| \geq 1$. Indeed, as $\psi\leq (4\pi)^{-1} r^{-1}$
where $r \geq 1$ by virtue of (18), a version of Holder's
inequality gives the bound
\begin{align*}
\int_{r\geq 1} |x - (\cdot)|^{-1} |a| \psi^2 
&\leq \left(\int_{r\geq 1} |a|^2 \psi\right)^{1/2} 
\left(\int_{r\geq 1} |x - (\cdot)|^{-2} \psi^3\right)^{1/2}\tag{56}  \\
&\leq (4\pi)^{-3/2}  |x|^{-1} (\ln|x|)^{1/2} .
\end{align*}
In particular, as $|x - (\cdot)|^{-1} |a| \psi^2$ is integrable 
for any $x \in\bR^3$, and $|a|/r$ is square integrable, the  
solution a to the second equation in (9) is given by 
$$
a|_x = (4\pi)^{-1} g^4 \int_{r\geq 1} |x - (\cdot)|^{-1} a \psi^2.
\eqno(57)
$$
Note that, (56) and (57) imply that $|\al| =r|a|\leq\ze g^4(\ln r)^{1/2}$
with $\ze$ a constant that is independent of  $\al$.
	
\medskip\noindent
{\it Step} 4. \   
To remove the factor of $(\ln r)^{1/2}$ from this estimate, first fix
$\de\in(0, 1/4)$; its value is  
determined by the chosen $\ve$.  Next, write the integral in (57) as a sum
of two parts, $s_+ + s_-$   where $s_+$  
is given by (57) with the integration domain now the region where $r
\geq\de |x|$; correspondingly, $s_-$ is  
given by (57) with the integration domain the ball of radius $\de |x|$
centered at the origin.  Note that  
$|s_+|$ is bounded by
\begin{align*}
& g^4 \ze \left(\int_{r>\de|x|} |a|^2 \psi\right)^{1/2} 
\left(\int_{r>\de|x|} |x - (\cdot)|^{-2} r^{-3}\right)^{1/2} \tag{58}\\
& \leq \ze_+ \si(|x|) |x|^{-1} (\ln \de)^{1/2} , 
\end{align*}
where $\ze_+$ is independent of $r, \de, \al$ and $g$, 
while $\si(s)^2 \equiv\int_{r>\de s} |a|^2 \psi.$
In particular, note that 
$$
 \lim_{s\ra\iy} \si(s) = 0. \eqno(59)
$$
	
Meanwhile, the necessary bound for $s_-$ requires two preliminary
observations.  The first is  that
$$
\int_{r<R} a \psi^2 = 0 \eqno(60)
$$
for any $R \geq 1$.  Indeed, this follows from symmetry considerations
after writing $a = \al \sin \ta d\va$ with  
respect to the Cartesian differentials 
$\{dx, dy, dz\}$ as $a = \al(r, \ta) (\cos \va dy - \sin \va dx)$.  The second 
key observation is that when $|y| \leq 2^{-1} |x|$, then
$$
|x - y|^{-1} = |x|^{-1} + |x|^{-2} \wp_1(x, y) ,\eqno(61)
$$
where $|\wp(x, y)| \leq\ze |y|$.  

Together, (60) and (61) imply that
$$
a_- =(4\pi)^{-1}g^4|x|^{-2}\int_{r\leq\de|x|}\wp_1 a\psi^2 .\eqno(62)
$$
In particular, (62) implies that
$$
|a_-| \leq \ze |x|^{-2} 
\left(\int |a|^2 \psi\right)^{1/2} 
\left(\int_{r\leq\de|x|} r^{-1} d^3y\right)^{1/2} 
\leq \ze_- \de |x|^{-1} ,\eqno(63)
$$
where $\ze_-$ is independent of both $|x|$ and $\de$.
	
The assertion of Proposition 5 follows directly from (58), (59) and
(63) by first choosing $\de = 2^{-1} (1 + \ze_-)^{-1} \ve$ 
to make $|a-_| < 2^{-1} \ve |x|^{-1}$; and then choose $|x|$
large so that $\si(|x|)$ in (59) is smaller  
than $2^{-1} (1 + \ze_+)^{-1} \ve/\ln(\de)^{1/2}$.

\section*{f) Some refined conclusions about $\psi$\\ at large $r$}

The purpose of this step is to refine the large $r$ bound of $\psi$ from
Proposition 4.  Indeed,  consider:

\begin{proposition}		
Suppose that $\al$ is a minimizer of a  $g > (6\pi)^{1/2}$ 
version of $\ce_0$.  Then the corresponding $\psi$ obeys  
$\psi = \psi_0 r^{-1} + o(r^{-1})$ at large $r$ 
where $\psi_0$ is a constant that is no larger 
than $\surd 2 g^{-2}$. 
\end{proposition} 

The remainder of this section is occupied with the

\medskip\noindent
{\it Proof of Proposition} 6. \   
The proof is divided into 3 steps.

\medskip\noindent {\it Step} 1. \ 
This step establishes that $\psi =
\psi_0 r^{-1} + o(r^{-1})$ with $\psi_0$ a constant.  For this
purpose, note that it is an immediate consequence of (8) that such
will be the case provided that
$$
\int |x - y|^{-1} |a|^2 \psi d^3y  = m_0 r^{-1} + o(r^{-1}) \eqno(64)
$$
for $r$ large, where $m_0$ is a constant.  And, this last conclusion,
follows directly from (18) and  
Proposition 5.  

\medskip\noindent	
{\it Step} 2. \  
This step constitutes a digression to establish

\begin{lemma}		
Suppose that $\al$ minimizes $a$ $g > 0$ version of $\ce_0$.  
Then either $\al$ is non-negative or non-positive. 
\end{lemma} 

\noindent
{\it Proof of Lemma} 7. \  
Since the $\al$ and $|\al|$ versions of (8) are identical, 
the corresponding solutions agree, and $\ce_0(|\al|) = \ce_0(\al)$.  
Thus, $|\al|$ minimizes $\ce_0$ if $\al$ does.  As $|\al|
= \al$ where $\al\geq 0$, it follows  from the unique 
continuation theorem of Aronszajn \cite{A} that $\al = |\al|$
everywhere if $\al$ is anywhere larger than zero.

With this lemma understood, the on going (and usually implicit)
assumption in the remainder of this article is that $\al\geq 0$.

\medskip\noindent
{\it Step} 3. \  Now, introduce	
$$
f(r) \equiv \int_{0\leq\ta\leq\pi}\al(r,\ta)\sin^2 \ta d\ta.\eqno(65)
$$
As $\al\geq 0$, so $f\geq 0$.  This function $f$ obeys the following
differential equation: 
$$
- f_{rr}+2r^{-2}f=g^4\int_{0\leq\ta\leq\pi}\psi^2\al\sin^2\ta d\ta,
\eqno(66)
$$
as can be seen by integrating both sides of the equation in the second
point of (9) using the measure $\sin^2 \ta d\ta$ on $[0, \pi]$.  
	
Using the fact that $r \psi$ has limit $\psi_0$ as $r \ra\iy$, 
this last equation implies that 
$$
-f_{rr} + 2 r^{_2} f > 2 r^{-2} f \eqno(67)
$$
at large $r$ if $\psi_0 > surd 2 g^{-2}$.  This last equation implies
that $f_{rr}< 0$ for large $r$.  Now, were $f_r$ ever negative, then
$f_r$ would become more negative as $r$ increased and thus the
condition $f > 0$ would be violated.  Hence, $f_r > 0$ for all
sufficiently large $r$.  Of course, this implies that $f$ is
increasing as $r$ tends to $\iy$, a conclusion which is forbidden by
Proposition 5.  Thus, $\psi_0 \leq \surd 2 g^{-2}$ as claimed.

\section*{g)  The behavior of $\al$ at large $r$}

Take $g > (6\pi)^{1/2}$ in this section.  This done, reintroduce
the constant $\psi_0$ from Proposition 6  and agree to write 
the latter as $\surd 2 g^{-2} e_0$ where $e_0 \in [0, 1]$.  It
then follows from (66) using the  
maximum principle that the function $f$ in (65) obeys
$$
f > r^{-p} \eqno(68)
$$
for any $p > p_0 \equiv 2^{-1} [(1 + 8 (1 - e_0^2))^{1/2} - 1]$.  
In fact, one can show without much difficulty that 
there is a constant $c_0 > 0$ such that
$$
\al = c_0 \sin \ta r^{-p_0} + o(r^{-p_0})  \eqno(69)
$$
as $r$ gets large.  With regard to (68) and (69), note that when $e_0
> 0,$ these equations imply that the magnetic field, $B = *da$, falls
off at an anomolously slow rate; as the classical dipole field
requires $\al = \co(r^{-1})$ at large $r$.  In any event, as is argued
momentarily $e_0 > 0$ as asserted in the third point of Theorem 2.
	
The claim that $e_0 > 0$ follows from the existence of a bounded function
$\si = \si(r, \ta)$ that  obeys the four conditions
\begin{itemize}
\item
$-\De\si + r^{-2} \al^2 \si = 0 $ \hfill (70)
\item
$\lim_{r\ra\iy}\si$ exists and equals 1.
\item
$\si > 0$ everywhere. 
\item
$\si(0) = \surd 2 g^{-2} e_0$.
\end{itemize}

To belabor the obvious, the last two lines of (70) are compatible only
if $e_0 > 0$.  

To consider the existence issue for $\si$, fix a smooth
function $\be: [0, \iy) \ra [0, 1]$ which vanishes on $[2, \iy)$, equals one
on $[0, 1]$ and is non-increasing.  Given $R > 1$, promote $\be$ to a smooth
function, $\be_R$, on $\bR^3$ by setting $\be_R(x) \equiv \be(|x|/R)$.
This done, consider solving for a function $\si_R$ that obeys
\begin{itemize}
\item
$-\De\si_R + \be_R r^{-2} \al^2 \si_R = 0$. \hfill (71)
\item
$\lim_{r\ra\iy} \si_R$ exists and equals 1.
\end{itemize}

The existence of a unique such function is not hard to establish and
the latter task is left to the  
reader.  Here are some of the properties of $\si_R$:  First, the maximum
principle guarantees that $\si_R > 0$  and, as $\al$ 
is not zero identically, that $\si_R < 1$.  Second, $\si_R \leq 1
- \co(r^{-1})$.   Third, $|\na\si_R| = \co(r^{-2})$.   
Indeed, these last two properties are consequence of the integral
equation equivalent of (71): 
$$
\si_R(x) = 1 - (4\pi)^{-1} \int |x - (\cdot)|^{-1} \be_R r^{-2} \al^2 \si_R.
\eqno(72)
$$
The maximum principle also guarantees that $\si_R < \si_{R'}$ when $R
> R'$.  This understood, it follows that the sequence
$\{\si_R\}_{R\ra\iy}$ is decreasing pointwise and so there is a unique
limit, $\si$.  Moreover, by virtue of the first line in (71) and the
bound $\si_R<1$ the functions in the set $\{\si_R\}$ are uniformly
continous with bounded first derivatives on any given compact set.
Moreover, they have uniformly continuous derivatives to any order on
any compact set that avoids the unit sphere.  Thus, the limit function
$\si$ is smooth where $r \geq 1$ and obeys the equation in the first line of
(70).  Also, $\si < 1$ everywhere and $\si\geq 0$ everywhere with equality
only if $\si\equiv 0$.

Can $\si$ vanish identically?  To prove that $\si > 0$ in the case $e_0 < 1$,
note first that under this assumption, 
$r^{-2} \al^2 \leq c \be_1 r^{-2-2p_0}$ with $p_0 > 0$ as
in (69) and with $c > 0$ a constant.  This understood, then (72) implies
the existence of some constant $c'$, independent of $R$, and such that
$\si_R \geq 1  c'$.  As $\si_R$ converges pointwise to $\si$, this last
equation implies that $\si > 0$ somewhere and hence everywhere on $\bR^3$ when
$e_0 < 1$.
	
In the case where $e_0 = 1$, a non-zero lower bound for $\si$ can still be
deduced from (72), albeit  
with more effort.  The start of this task uses the fact that
$\si_R$ converges pointwise to $\si$ to deduce  
from (72) that
$$
\si(x) \geq 1 - (4\pi)^{-1} \int |x - (\cdot)|^{-1} r^{-2} \al^2.\eqno(73)
$$
Now, to estimate the integral, consider breaking the integration
domain into three regions: Region 1 has $r > 2 |x|$, Region 2 has
$|x|/2 \leq r \leq 2 |x|$ and Region 3 has $r \leq |x|/2$.

The contribution to the integral in (73) from Region 1 is no more than
$$
(4\pi)^{-1} \int_{r>2|x|} r^{-3} \al^2 \leq \ze \int_{r>2|x|} |a|^2 \psi,
\eqno(74)
$$
where $\ze$ is an $x$-independent constant.  As the integral of $|a|^2
\psi$ is finite (as asserted by (55)), so the function of $|x|$
defined by the right hand integral in (74) tends uniformly to zero as
$|x|$ tends to infinity.  Meanwhile, the contribution to the integral
on the right hand side of (73) from Region 2 is no greater than
\begin{align*}
& (4\pi)^{-1} \left(\sup_{|x|/2\leq r\leq 2|x|} \al^2(y)\right) 
\int_{|x|/2\leq r\leq 2|x|} |x - (\cdot)|^{-1} r^{-2} \tag{75}\\
& \leq \ze \sup_{|x|/2\leq r\leq 2|x|} \al^2(y) ,
\end{align*}
where $\ze$ is, again, and $x$-independent constant.  Note Proposition
5 asserts that the right hand side of (75) tends uniformly to zero as
$|x|$ tends to infinity.
	
Finally, consider the contribution to the integral on the right hand
side of (73) from Region 3.  This contribution is no greater than
$$
c |x|^{-1} \int_{r<|x|} r^{-2} \al^2 .\eqno(76)
$$
To bound the latter expression, divide the region of integration into
the domains $\{A_n: 1 \leq n \leq |x|\}$ where the index $n$ is an
integer and $A_n = \{y: n \leq |y| \leq n + 1\}$.  Meanwhile, let sn
denote the integral of $r^{-3} \al^2$ over $A_n$.  In this regard, note
that $r^{-3} \al^2$ is integrable over all of $\bR^3$ by virtue of
(55), and so $\sum_{1\leq n<\iy} sn < \iy$.  Let $S$ denote the value
of this infinite sum.  It then follows that the expression in (76) is
no greater than
$$
c' |x|^{-1} \sum_{1\leq n\leq |x|} n s_n .\eqno(77)
$$
Now, as the $\sum_{1\leq n<\iy} s_n$ is finite, given $\ve$, there
exists $N$ such that $\sum_{n\geq N} s_n < \ve$.  This understood,
then (77) is no greater than
$$
\ze\left(\sum_{n\geq N}s_n+N|x|^{-1}\sum_{1\leq n\leq N}s_n\right) 
\leq \ze' (\ve + N |x|^{-1}) .  \eqno(78)
$$
Here, both $\ze$ and $\ze'$ are $x$-independent constants.  This last
bound implies that Region 3's contribution to the integral on the
right hand side of (73) tends uniformly to zero as $|x|$ tends to
infinity.
	
Thus, the analysis for the three regions has established the
following:  Give $\ve > 0$, there  
exists $r_{\ve}> 1$ such that 
$$
\si(x) \geq 1 - \ve\quad\text{ when }\;  |x| > r_{\ve} .\eqno(79)
$$
This last inequality establishes that $\si > 0$ even when $e_0 = 1.$
By the way, this last inequality also establishes that $\si$ tends
uniformly to 1 as $r$ tends to infinity.

Now turn to the final point in (70).  To establish this point,
multiply both sides of the equation $-\De\psi + r^{-2} \al^2 \psi =
\rho$ by $\si_R$ and integrate the resulting equation over $\bR^3$.
Integrate by parts twice to place the Laplacian on $\si_R$.  Then,
invoke (71) to obtain the equality
$$
4\pi\surd 2g^{-2}e_0+\int(1-\be_R)r^{-2}\al^2\psi\si_R= \si_R(0).
\eqno(80)
$$
Here, the fact that $\si_R$ is harmonic in the ball has been used to
identify $\int\rho\si_R$ with $\si_R(0)$.  In any event, this last
equation asserts that $\si_R(0) > 4\pi \surd 2 g^{-2} e_0$ for all
$R$, and so $\si(0) > 0$ for all $R$.  This proves that $\si$ is not
identically zero when $e_0 > 0$ and thus when $e_0 = 1$.
	
Equation (80) implies the final line of (70) by virtue of (55), for
the latter implies that the integral term on the left side of (80)
converges uniformly to zero as $R \ra\iy$.

\section*{h)  A lower bound for the magnetic energy}

The purpose of this subsection is to derive a lower bound for $\int
|\na a|^2$. Here is the precise statement:

\begin{proposition}		
There exists a constant $c \geq 1$ with the following significance:  Fix
$g > (6\pi)^{1/2}$ and  let $\al$ be a minimizer 
of the $g$ version of $\ce_0$ in $(10)$.  Set $a = \al \sin\ta
d\va$.  Let $U$ denote the annulus  
where $1 \leq r \leq 1 + c g^{-1}$.  Then
$$
g^{-2} \int_U |\na a|^2 \geq  c^{-1} g .\eqno(81)
$$
\end{proposition}

The remainder of this section is occupied with the

\medskip\noindent
{\it Proof of Proposition} 8. \  
The proof is divided into four steps.

\medskip\noindent	
{\it Step} 1. \  
First, let $c_1$ denote the constant which appears in Proposition 4, and
introduce $\cC$ to  
denote the set of Sobolev class $L^2_1$ functions $u$ on $\bR^3$ which vanish
where $r \geq 1 + 64\pi c_1 g^{-1}$.  The  
claim here is that if $g \geq 64\pi c_1$, then there exists a unique
$\ka\equiv \ka(\al) \in (0, \iy)$ with the property that    
\begin{align*}
&3 (8\pi)^{-1} \int_{r\leq 1} \psi\tag{82}\\
& = \sup_{u\in\cC} \left\{ 3 (4\pi)^{-1} \int_{r\leq 1} u  
- 2^{-1} \left( \int |\na u|^2 +  \ka^2 g^2 \int_{r\geq 1} 
r^{-2} u^2\right) \right\} .
\end{align*}
	
To see that $\ka$ exists, remark first that for any choice of $\ka$,
the supremum on the right hand side of (82) is achieved by a unique
function, $u_{\ka}$, which can be written down in closed form.  To
write $u_{\ka}$, first introduce
\begin{itemize}
\item
$d \equiv 64\pi c^1 g^{-1}$. \hfill (83)
\item
$p \equiv 2^{-1} ( (1 + 4 \ka^2 g^2)^{1/2} + 1)$.
\item
$p' \equiv 2^{-1} ((1 + 4 \ka^2 g^2)^{1/2} - 1)$.
\end{itemize}

With these definitions understood, here is $u_{\ka}$:
\pagebreak

\mbox{}\hfill (84)
\begin{itemize}
\item
$u_{\ka} = (8\pi)^{-1} (1 - r^2 + [(1 + d)^{p+p'} - 1]/
[p (1 + d)^{p+p'} + p'])$   	where $r \leq 1$. 
\item
$u_{\ka} = (4\pi)^{-1} [(1 + d)^{p+p'} r^{-p} - r^{p'}]/
[p (1 + d)^{p+p'} + p']$ where $1 \leq r \leq 1 + d$.
\end{itemize}

Since the supremum on the right hand side of (82) is $3 (8\pi)^{-1}
\int_{r\leq 1} u_{\ka}$, integration of the first line in  
(82) finds the supremum on the right hand side of (82) equal to 
$$
f(\ka) \equiv (40\pi)^{-1} + (16\pi)^{-1} 
[(1 + d)^{p+p'} - 1] [p (1 + d)^{p+p'} + p']^{-1} .\eqno(85)
$$
	
Now, there are four key observations about $f(\cdot)$ in (85):  First,
$f$ is a continuous function of  $\ka\in [0, \iy)$.  
Second, $f$ is a monotonically decreasing function on $[0, \iy)$.
Third, $\lim_{\ka\ra\iy}f(\ka) =  (40\pi)^{-1}$.  Finally, 
$$
f(0) = (40\pi)^{-1} + (16\pi)^{-1} (1 + (64\pi c_1)^{-1} g )^{-1}. 
\eqno(86)
$$
If the right hand side of (86) is greater than 
$3 (8\pi)^{-1} \int_{r\leq 1} \psi$, then it \linebreak
follows from (13) and the first  two points of Proposition 4 that \linebreak
$3 (8\pi)^{-1} \int_{r\leq 1} \psi$
lies in the range of $f$ and so there exists some $\ka\in (0, \iy)$ 
which makes (82) true.  Moreover, there will be a unique
such $\ka$ since $f$ is monotonically  
decreasing.  
	
Now, due to (13) and Proposition 4, the right hand side of (86) can be
guaranteed greater than $3 (8\pi)^{-1} \int_{r\leq 1} \psi$ when 
$$
(16\pi)^{-1} (1 + (64\pi c_1)^{-1} g)^{-1} \geq 2 c_1/g ,\eqno(87)
$$
which occurs when $g \geq 64\pi c_1$.
	
\medskip\noindent
{\it Step} 2. \   Here is the next point:
$$
\text{There exists $K \geq 1$ such that if $g \geq 64\pi c_1$, 
then $\ka(\al) \geq 1/K$.}\eqno(88)
$$
Indeed, suppose that $\ka = 10^{-5} (64 \pi c_1)^{-1}$.  Also, assume that $g >
10^{10}64\pi c_1.$ Thus $\ka g > 10^5$.  With  
this understood, $p, p' = \ka g + o(10^{-5})$ and $(1 + d)^{p+p'} = 1 +
(128\pi c_1) \ka + o(10^{-5})$.  This implies that  
$$
f(\ka) \geq (40\pi)^{-1} + (32\pi)^{-1} (128\pi c_1 \ka)/
(\ka g + \ka g)^{-1} = (40\pi)^{-1} + 2 c_1 g^{-1} .\eqno(89)
$$
Because of (13) and the assertions of Proposition 4, the right hand
side of (89) can not be smaller than $3 (8\pi)^{-1} \int_{r\leq 1}
\psi$.  Thus, since $f$ is monotonically decreasing, it follows that
$\ka(\al)$ in (82) is larger than $10^{-5} (64\pi c1)^{-1}$ when $g >
10^{10} (64 \pi c_1)$.  This fact implies the existence of $K$ which
makes (88) hold for the larger range of $g$.

\medskip\noindent
Step 3:  Now, remember that  
$$
3 (8\pi)^{-1} \int_{r\leq 1} \psi
= sup_u \left\{3 (4\pi)^{-1} \int_{r\leq 1} u 
- 2^{-1} \int (|\na u|^2 + |a|^2 u^2) \right\} ,\eqno(90)
$$
where the supremum on the right hand side is taken over the set of
functions $u$ on $\bR^3$ with both $\na u$  
and $r^{-1} u$ square integrable.  
Note that $u_{\ka}$ is such a function for any $\ka$.
In particular, take 
$$\ka =  \min(1/K, 10^{-5} (64\pi c_1)^{-1}).$$  
In this case, the fact that $f(\ka)$ is decreasing, (82) and (90) imply that
$$
\int|a|^2 u_{\ka}^2\geq\ka^2 g^2\int_{r\geq 1}r^{-2}u_{\ka}^2. \eqno(91)
$$
This last inequality with the second point in (84) imply that there is
a constant $C \geq 1$ such that when  
$g \geq C$, then
$$
\int_U |a|^2 \geq C^{-2} g .\eqno(93)
$$
	
Since $a|_{r=1} = 0$, this last inequality implies the existence of
the constant c which makes (81) hold.

\section*{i) The value of $e_0$ and the form of $\psi$ 
and $\al$ where $r \geq 10$}

The previous sections found a constant $e_0 \in (0, 1]$ that controls
the large $r$ asymptotics of both $\psi$ and $\al$ in as much as $\psi
\sim \surd 2 g^{-2} e_0 r^{-1}$ and $ \al\sim c_0 \sin
r^{-[(9-8e_0^2)^{1/2}-1]/2}$ as $r \ra\iy$.  In particular, these
observations imply that $\psi\leq \surd 2 g^{-2} e_1 r^{-1}$ and
$\al\geq c_1 \sin \ta r^{-1}$ at large $r$ with $e_1$ and $c_1$
positive constants.  This understood, the purpose of this section is
to bound e1 from above and $c_1$ from below and to establish a lower
bound for the radius r where this upper bound for $\psi$ and lower
bound for $\al$ are reasonable.  The fruits of this labor are
summarized by

\begin{proposition}			
There exists $\ga\geq 1$ with the following significance:  
Suppose that $g > (6\pi)^{1/2}$ and 
that $\al$ minimizes the $g$-version of $\ce_0$.  
At points where $r \geq \ga$, this $\al$ and its associated $\psi$ obey
\begin{itemize}
\item
$\psi\leq\surd 2 g^{-2}e^{-g^{1/4}/\ga} r^{-1}$.
\item
$\al\geq\ga^{-1} g^{1/2} \sin \ta r^{-1}$ .
\end{itemize}
\end{proposition}

The remainder of this section is dedicated to the 

\medskip\noindent
{\it Proof of Proposition} 9. \  
As the proof is long, it is broken into
twelve steps.  Take $g > (6\pi)^{1/2}$ in all  steps below.

\medskip\noindent
{\it Step} 1. \   
According to (93), there is a positive, $g$-independent constant $c$ such
that $|a|^2$ has  integral greater than 
$c^{-1} g$ over the region where $r < 1 + c/g$.  This
step provides a refinement with a  
proof that the mass of this integral can not concentrate where sin $\ta$
is zero.  Here is the precise statement: 

\begin{lemma}		
There exists $c > 1$ such that when $g > c,$ then
$$
\int_{r\leq 1+c/g,\sin\ta>1/c} |a|^2 \geq c^{-1} g .\eqno(94)
$$
\end{lemma}

\medskip\noindent
{\it Proof of Lemma} 10. \   
Suppose that the lemma were false.  
Then, given $R \geq 1$ and $\ve > 0,$ there 
would exist arbitrarily large values for $g$ for which 
$$
\int_{r\leq 1+R/g, \sin\ta>1/R^2} |a|^2 \leq\ve g/R .\eqno(95)
$$
Given that such is the case, fix $\de\in (0, 1/1000)$ and let
$u_{\de}$ denote the following function:
$$
u_{\de}\equiv \max(0, 3 (8\pi)^{-1} [(1 - r^2) 
+ \de (2 + (1 - 3 \cos^2 \ta) r^2))].\eqno(96)
$$
Note that $u_{\de}  < \de$ where $r \geq 1$ and 
$u_{\de} = 0$ where $r > 1 + 2 \de$.  Moreover,
$$
u_{\de}< 9 (8\pi)^{-1} \de\sin^2 \ta     \eqno(97)
$$
where $r \geq 1$ and $\sin^2 \ta < 2/3$.  
	
The preceding properties of $u_{\de}$ imply that
\begin{align*}
& \int\rho u_{\de}-2^{-1}\int(|\na u_{\de}|^2+|a|^2 u_{\de}^2)\tag{98}\\ 
& \geq(40 \pi)^{-1} + 2 \de - c_1 \de^2 \\
&\quad-  \de^2\left(\int_{r\leq 1+R/g,\sin\ta>1/R^2}|a|^2 
+ R^{-8} \int_{r<1+R/g} |a|^2\right) .
\end{align*}
With (98) understood, take $\de = R/(2g)$.  This done, the right hand side
of (98) is no smaller than 
$$
(40 \pi)^{-1} + R/g (1 - \ze (\ve + R^{-4} + R/g)) ,\eqno(99)
$$
where $\ze$ is a $g$-independent constant.  Indeed, (99) follows
immediately from (98) and (95) given that $\int_{r<1+R/g} |a|^2$ is
bounded by a \linebreak
$g$-independent multiple of $R^2 g$.  In this regard, the
latter bound \linebreak
is obtained as follows: Note first that $\int_{r<r+R/g}
|a|^2$ is bounded by \linebreak
$(R/g)^2 \int_{r<1+R/g}|\na a|^2$ since $a = 0$
where $r = 1$.  Meanwhile, according to (14), the integral of $|\na
a|^2$ over the whole of $\bR^3$ is equal to $g^4$ times that of $|a|^2
|\psi|2$, and the latter, by virtue of Proposition 4, is no greater
than $c_1 g^{-1}$.
	
Now, the point is that when $R$ is large, the lower bound in (99) is
not compatible with Proposition 4 because $\psi$ and not $u_{R/(2g)}$
maximizes the expression $\int\rho u d^3x - 2^{-1} \int (|\na u|^2 +
|a|^2 u^2) $ as $u$ ranges over the smooth functions on $\bR^3$ for
which $|u|/r$ is square integrable.

\medskip\noindent
{\it Step} 2. \   
This step proves that there is a $g$-independent constant $\ul{c} > 1$ and, for
each $g > \ul{c}$,  there is a number 
$r_0 \in (1, 1 + \ul{c}/g)$ such that the function $f$ in (65) obeys
$$
f(r_0) \geq \ul{c}^{-1} g^{1/2} .\eqno(100)
$$
In this regard, note that (100) and (66) imply that
$$
f(r) \geq \ul{c}^{-1} g^{1/2} r^{-1} \eqno(101)
$$
for all $r \geq r_0$.
	
The proof for existence of such constant $\ul{c}$ and $r_0$ uses three
previously established facts: The first fact is the assertion in Lemma
10.  The second, a converse of sorts, was derived to end the proof of
Lemma 10: The constant $c$ in Lemma 10 can be chosen so that
$\int_{1\leq r\leq 1+c/g} |a|^2 \leq c g$.  
The third, a consequence of (9) and Proposition 4, 
is that $c$ can be chosen so that $\int_{1<r<1+c/g} |\na a|^2 < c
g^3$.  By virtue of these three facts, there exists a $g$-independent
constant $c' \geq 1$ and a possibly $g$-dependent number $r_0 \in (1, 1 +
c'/g)$ such that $m \equiv |a|\,\big|_{r=r_0}$ obeys
\begin{itemize}
\item
$c^{\prime -1} g \leq \int_{\sin\ta>1/c'} m^2 d\ta \leq c'
g^2$. \hfill (102)
\item
$\int_{\sin\ta>1/c'} m_{\ta}^2 d\ta \leq c' g^4$.
\end{itemize}
As argued momentarilly, these inequalities imply that
$$
\int_{\sin\ta>1/c'}m d\ta\geq 2^{-3/4}c^{\prime -3/2}g^{1/2}. \eqno(103)
$$
The desired inequality in (100) follows from (103) 
with $\ul{c} = 2^{-3/4} c^{\prime -7/2}$.

To establish (103), note first that $\int_{\sin\ta>1/c'} m^2 d\ta \leq
\hat{m}\int_{\sin\ta>1/c'} m d\ta$ where $\hat{m}$ denotes the maximum
value of $m$ where $sin \ta > 1/c'$.  Thus, the left most inequality
in the first point of (102) requires
$$
c^{\prime -1}\hat{m}^{-1} g^2 \leq \int_{\sin\ta>1/c'} m d\ta .\eqno(104)
$$
To obtain the requisite upper bound on  $\hat{m}$, 
first note that when $m(\ta) =\hat{m}$, 
then $m(\ta') \geq 2^{-1}$    provided that 
$$
|\ta - \ta'| < 4^{-1} c^{\prime -1} g^{-4} \hat{m}^2 .\eqno(105)
$$
Indeed, this follows from the bottom point in (102). Second, observe
that (105) is consistent with  
the right hand inequality in the top point of (102) only if
$$
2^{-1}\hat{m}^2 (4^{-1} c^{\prime -1} g^{-4}) \hat{m}^2 
\leq c' g^2 .\eqno(106)
$$
Thus, $\hat{m}\leq (8 c^{\prime 2})^{1/4} g^{3/2}$.  
Insert this last bound into (104) to obtain (103).   

\medskip\noindent
{\it Step} 3. \   
The step constitutes a digression of sorts to state and then prove

\begin{lemma}			
Let $\be$ be a function of the spherical coordinates $(r, \ta)$ that
vanishes where $r \leq 1,$ has limit zero as $r \ra\iy$ and such that
$r^{-2} \be, r^{-1} \be_r$ and $r^{-2} (\sin \ta)^{-1} (\sin
\ta\be)_{\ta}$ are all square integrable.  Then
$$
\int r^{-2}(\be_r^2 + r^{-2} (\sin \ta)^{-2} (\sin \ta\be)_{\ta}^2) 
\geq g^4 \int r^{-2} \be^2 \psi^2\eqno(107)
$$
with equality if and only if $\be$ is a multiple of $\al$.
\end{lemma}

\noindent
{\it Proof of Lemma} 11. \    
Suppose that (107) is violated by some non-trivial $\be$.  
Then, as is explained 
momentarily, there exists a non-negative violator, $\be$, with 
$$
- \be_{rr}- r^{-2} ((\sin \ta)^{-1} (\sin \ta\be)_{\ta})_{\ta} 
- g^4 \psi^2 \be \leq 0 \eqno(108)
$$
where $r \geq 1$ and with a strict inequality on some open set.  Take
this last equation and multiply by $\al$, then integrate the resulting
inequality over the $r \geq 1$ portion of $\bR^3$.  As $\al > 0$, the
result is a negative number.  However, as $\al$ also obeys (9), two
applications of integration by parts contradict this last assertion.
	
To prove (108), note first that if there exists $\be$ that violates (107),
there exists such a function $\be$ which is positive and has compact
support in some very large radius ball.  This said, fix $R \geq 1$ and
let $B_R \subset \bR^3$ denote the ball of radius $R$.  
When $R$ is large, standard potential theory finds 
a unique function $\be_R$ on $B_R-B_1$, vanishing on the
boundary of this domain, positive on its interior with maximum 1 and
satisfying
$$
- (\be_R)_{rr}- r^{-2} ((\sin \ta)^{-1} 
(\sin \ta\be_R)_{\ta})_{\ta} - g^4 \psi^2 \be_R = - \la_R \be_R    
\quad\text{where }\;r \geq 1\eqno(109)
$$
with $\la_R > 0$.  Potential theory can also be used to prove that $\la_R$
increases with $R$.  Meanwhile, view (109) at a local maximum of $\be_R$ to
see that $\la_R \leq g^4 \max_{r\geq 1} \psi^2$.
	
Now, consider that $\psi = \surd 2 g^{-2} e_0 r^{-1} + o(r^{-1})$.
This understood, it follows from (109) that there exists $r_0 \geq 1$
such that when $R$ is large, then $\be_R$ has no local maxima where $r
\geq r_0$.  In fact, this last equation implies that $\be_R \leq
e^{-\la_Rr/2}$ where $r \geq r_0$.  In any event, as $\be_R = 1$ at
some $r < r_0$ for all large $R$, and as $\la_R$ is increasing and
bounded, (109) coupled with standard elliptic regularity theory finds
that $\lim{R\ra\iy} \be_R \equiv \be$ exists and has the following
properties: First, $\be$ is a smooth function where $r > 1$ and
vanishes at $r = 1$.  Second, $\be$ has maximum 1.  Third, all of the
following are square integrable: $r^{-1} \be, r^{-2} (\sin \ta)^{-1}
(\sin \ta\be)_{\ta}$ and $r^{-2} \be$.  Finally, $\be$ is a violator
of (107) and obeys (108) with the inequality holding on a non-empty
set.  (In fact, $\be$ obeys (109) with $\la_R$ replaced by $\la\equiv
\lim_{R\ra\iy} \la_R.$)

\medskip\noindent
{\it Step} 4. \   
This step uses (107) to obtain an upper bound on the size of
$\psi$ where $r \geq 4$.  The  precise statement is

\begin{lemma}			
There exists a $g$-independent constant $\xi$ with the following
significance:  At points where $r \geq 4,$ 
the function $\psi$ obeys  $\psi\leq\xi g^{-2} r^{-1}.$
\end{lemma}

\noindent
{\it Proof of Lemma} 12. \  
To begin, take $\be$ in (107) as follows:
\begin{itemize}
\item
$\be(r, \ta) = (r - 1) \sin^{1/4} \ta$   where  $r \leq 2$.\hfill (110)
\item
$\be(r, \ta) = 4 r^{-2} \sin^{1/4}\ta$   where $r \geq 2$.
\end{itemize}
The left hand side of (107) for this choice of $\be$ is finite, some
constant.  This understood, then  
(107) implies the existence of a $g$-independent constant $\xi_1$ such that
$$
\int_{2\leq r\leq 8} \psi^2 \sin^{1/2}\ta \leq g^{-4} \xi_1 .\eqno(111)
$$
	
Next, use the fact that $-\De\psi\leq 0$ where $r \geq 1$ and standard
Green's function techniques to  
find a $g$-independent constant such that 
$$
\psi\leq \ka r^{-1} \int_{2\leq r\leq 3} \psi .\eqno(112)
$$
at points where $r \geq 4$.  (Remember when deriving (112) that $\psi =
|\psi|$.)  With (111) and (112), the  lemma follows using Holder's 
inequality and the fact that $\sin^{-1/2}\ta$ is locally integrable on
$\bR^3$.

\medskip\noindent
{\it Step} 5. \   
This step uses the bound in Lemma 12 to obtain an upper bound on the integral 
over the region where $r \geq 8$ of $r^{-2} \al^2 \psi^2$.	
To obtain this bound, fix a function, $\chi$, of $r$ that equals 1 where $r
\geq 8$ and zero where $r \leq 4$.   
This done, multiply both sides of the equation in the first point of
(8) by $\chi\psi$ and then integrate the  result of $\bR^3$.  
Two applications of integration by parts and an appeal
to Lemma 12 yields an  inequality of the form
$$
\int_{8\leq r} (|\na\psi|^2 + r^{-2} \al^2 \psi^2) 
\leq\xi_1 \int_{4\leq r\leq 8} \psi^2 \leq \xi_2 g^{-4} ,
\eqno(113)
$$
where $\xi_1$ and $\xi_2$ are $g$-independent constants.

\medskip\noindent
{\it Step} 6. \   
To start this step, decompose $\al$ as $\al =\hat{\al}+\be$, 
where $\hat{\al}  =\al $ at $r = 8$ but otherwise  solves the equation
$$
-(\hat{\al}_{rr}+r^{-2}((\sin \ta)^{-1}(\sin\ta\hat{\al})_{\ta})_{\ta})= 0,
\eqno(114)
$$
while $\be = 0$ at $r = 8$ and obeys
$$
-(\be_{rr} + r^{-2} ((\sin \ta)^{-1} (\sin\ta\be)_{\ta})_{\ta})
= g^4 \psi^2 \al .\eqno(115)
$$
Both $\hat{\al}$  and $\be$ are solutions to their respective
equations on $\bR^3$ that limit to zero as $r \ra\iy$. In this  
regard, note that such a decomposition can be found using standard
properties of the Laplacian on  $\bR^3$ as  $\hat{\al}$ 
and $\be$ are obtained by first solving the equations 
\begin{itemize}
\item
$-\De\hat{a} = 0$   where $r \geq 8$ with $\hat{a}|_{r=16} =
a|_{r=16}$, \hfill (116)
\item
$-\De b = g^4 \psi^2 a$ where $r \geq 8$ with   $b|_{r=16} = 0$, 
\end{itemize}
and then writing $\hat{a} = \hat{\al}\sin\ta d\va$ and 
$b =\be \sin \ta d\va$. 
	
In any event, the purpose of this step is to obtain a pointwise bound
on $\be$.  In particular, as  $|b| = r^{-1} |\be|$, 
such a bound can be obtained using the Dirichelet
Green's function for $\De$ in  
conjunction with (116).  Indeed, this strategy finds
$$
|x|^{-1} |\be(x)| 
\leq (4\pi)^{-1} g^4 \int_{8\leq r} |x - (\cdot)|^{-1} r^{-1} \psi^2 \al.
\eqno(117)
$$
This last inequality understood, use Lemma 12 to eliminate one power
of $\psi$ and so bound the right side of (117)
\begin{align*}
& g^4 \int_{8\leq r} |x - (\cdot)|^{-1} r^{-1} \psi^2 \al\tag{118}\\  
& \leq\xi g^2 \int_{8\leq r} |x - (\cdot)|^{-1} r^{-2} \psi\al\\  
& \leq\xi g^2 \left(\int_{8\leq r} |x - (\cdot)|^{-2} 
r^{-2}\right)^{1/2} \left(\int_{8\leq r} r^{-2} \psi^2 \al^2\right)^{1/2}.
\end{align*}
Here, the left most inequality results from an application of Holder's
inequality.  Next, plug (113) into this last inequality to discover a
$g$-independent constant $\xi$ such that
$$
|\be| \leq \xi r^{1/2}\eqno(119)
$$
at points $x$ with $|x| = r$.

\medskip\noindent {\it Step} 7. \ This step studies the behavior of
$\hat{\al}$.  For this purpose, it proves useful to reintroduce the
function $f(r)$ from (65) and (101).  This done, the resulting
analysis of $\hat{\al}$ is then summarized by the following assertion:
There exists a $g$-independent constant $\xi$ such that
$$
|\hat{\al}|_{(r,\ta)} - r^{-1} f|_{r=16} \sin \ta\big|
\leq \xi r^{-2} f|_{r=16} \sin \ta   \eqno(120)
$$
where $r \geq 16$.  With regards to this last equation, note that the
maximum principle guarantees that is non-negative. 
	
To prove the assertion, and for use subsequently, it proves useful to
introduce the functions $\nu\equiv (r^2 \sin \ta)^{-1} \al$ and 
$\hat{\nu}\equiv (r^2 \sin \ta)^{-1}\hat{\al}$. Next, interpret and $\nu$ 
and $\hat{\nu}$  as functions on $\bR^5$ by writing 
$r$ and $\ta$ in terms of standard Cartesian coordinates as $r=(x_1^2+\cdots
 + x_5^2)^{1/2}$ and $\ta ={\rm Arccos}\,(x_5/r)$.   
This done, then the second line in (9) reads $-\De_5\nu- g^4 \psi^2
\nu$ and (114) becomes   
\begin{itemize}
\item
$-\De_5 \hat{\nu}= 0$ where $r > 8$. \hfill (121)
\item
$\hat{\nu}|_{r=8} = \nu|_{r=8}$.
\item
$\hat{\nu}\ra  0$ as $r \ra\iy$.  
\end{itemize}
Here, $\De_5$ denotes the standard Laplacian on $\bR^5$.  
All this understood, it follows from (121) using 
standard Green's function techniques that 
$$
\hat{\nu}\leq \xi_1 r^{-3}\int_0^{\pi}\nu_{r=8} \sin^3 \ta d\ta
=\xi_2 r^{-3} f_{r=8} \eqno(122)
$$
where $\xi_{1,2}$ are $g$-independent constants and where $f$ is the function
of $r$ from (65). 
	
Put this last bound in the bank temporarily and consider the expansion
of $\hat{\nu}$  as a sum of  spherical harmonics.  
That is, write $\hat{\nu}= \sum_{\la}\hat{\nu}^{\la}\ka^{\la}$
where $\ka^{\la}$ is a function only of $\ta$ and solves the  
eigenvalue equation 
$$
-(\sin \ta)^{-3}((\sin \ta)^3\ka^{\la}_{\ta})_{\ta}=\la\ka^{\la}.\eqno(123)
$$
In this regard, the lowest eigenvalue is $\la = 0$ with the constants as
eigenfunctions.  The next lowest  eigenvalue is $\la = 4$ 
with $\cos \ta$ as the eigenvalue.  Meanwhile, $\hat{\nu}^{\la}$ 
is a constant multiple of $r^{-(3+\sq{3+4\la})/2}$.  
This last point understood, then (120) follows directly.

\medskip\noindent
{\it Step} 8. \  
This step uses (119) and (120) to prove the first assertion
of Proposition 9.  For  
this purpose, note that these two inequalities directly imply the
existence of $g$-independent constants $\hat{m}\geq 1$
and $r_* \geq 16$ such that 
$$
\al\geq\hat{m}^{-1} g^{1/2} r^{-1} \sin \ta -\hat{m}r^{1/2}.\eqno(124)
$$
where $r \geq r_*$.  This understood, return now to the equation in the
first point of (8) for $\psi$.  Multiply  both sides of this equation 
by $\psi$ and then, at each fixed $r \geq r_*$,
integrate the result over the constant $r$  sphere to obtain 
the following equation for the function $q(r)\equiv \int_0^{\pi}
\psi^2|_r\sin \ta d\ta$: 
$$
-r^{-2} (r^2 q_r)_r + 2\int_0^{\pi} (\psi_r^2 + r^{-2} 
(\psi_{\ta}^2 + \al^2 \psi^2)) \sin \ta d\ta = 0 .\eqno(125)
$$
	
This last equation implies the inequality
$$
- r^{-2}(r_2 q_r)_r + r^{-2} \la(r) q \leq 0 ,\eqno(126)
$$
where $\la(r)$ is the smallest eigenvalue for the operator 
$$L|_r\equiv -(\sin \ta)^{-1}(\sin \ta (\cdot)_{\ta})_{\ta}+\al^2|_r.$$  
In this regard, use (124) with standard eigenvalue estimation techniques to
find a $g$-independent constant $\xi\geq 1$ such that
$$
\la(r) \geq \xi^{-1} g^{1/2} r^{-2} - \xi r.\eqno(127)
$$
Next, use (127) to find $g_1$ and a $g$-independent constant 
$m_* \geq 1$ such that $\la(r)\geq m_*^{-1} g^{1/2}$ when $g \geq 
g_1$ and $m_*\leq r \leq m_* + 1$.
	
Given this lower bound on $\la$ where $m_* \leq r \leq m_* + 1$ 
and Lemma 12's bound for $q$ at $r = m_*$, 
the differential inequality in (126) implies the existence of a
$g$-independent constant $\mu\geq 1$ such  
that when $g \geq g_1$, then
$$
q(r) \leq\mu g^{-4}e^{-g^{1/4}/\mu}.\eqno(128)
$$
where $r = m_* + 1$.  This last bound and the maximum principle then
bound $q(r)$ by $\mu g^{-4}$ at all $r \in [m_* + 1, m_* + 3]$.
Finally, with the latter bound in hand, the arguments from Step 4
provide a $g$-independent constant $\mu'\geq 1$ and the pointwise
bound of $\psi$ by $g^{-2}e^{-g^{1/4}/\mu'}r$ where $r \geq m_* + 2$.
This pointwise bound directly gives the first assertion of Proposition
9.

\medskip\noindent 
{\it Step} 9. \ 
This last step uses the first
assertion in Proposition 9 to obtain the second.  To begin,
reintroduce $\hat{\al}$ and $\be$ from Step 6.  Then, the following
assertion restates some facts from previous steps: There exists a
$g$-independent constants $\mu\geq 1$ and $\mu'$ such that when $g
\geq\mu$, then
\begin{itemize}
\item
$\psi\leq\mu g^{-2} r^{-1}e^{-g^{1/4}/\mu}$ \quad   	
at all points with $r \geq\mu$. \hfill (129)
\item
$\hat{\al}\geq\mu^{-1} g^{1/2} r^{-1} \sin \ta$ \quad  	
at all points with  $r \geq \mu$.
\item
$|\be|\leq\mu'$ \hspace*{25mm} at all points with $\mu\leq r \leq\mu + 1$.
\end{itemize}
This last point understood, introduce $\nu\equiv 
(r^2 \sin \ta)^{-1} \al$ and decompose
the latter where $r \geq\mu$ as the  
sum $\nu_0 + \nu_1$ where these new functions obey
\begin{itemize}
\item
$-\De_S \nu_0 = 0$ \hspace*{15mm} with $\nu_0|_{r=\mu} = \nu$. \hfill (130)
\item
$-\De_S \nu_1 = g^4 \psi^2 \nu$ \qquad with $\nu_1|_{r=\mu} = 0$.
\end{itemize}
To proceed, observe from (69) that $r^p\al$ is bounded on $\bR^3$ 
as long as $p\leq  p_0$ with $p_0$ 
as described just prior to (69).  Thus, said, then it
follows from (128) that with $\ve > 0$ specified, there exists $g_{\ve}$ such
that when $g > g_{\ve}$, then $r^{1-\ve}\al$ is bounded.  
This said, then $r^{3-\ve}\nu$ is bounded on $\bR^5$.  
Let $z(\nu) \equiv \sup_{r\geq\mu} (r^{3-\ve} \nu)$ and define $z(\nu_1)$
analogously.  Then, use the Green's function of $-\De_S$ to conclude
from the second line of (130) and the top line in (129)
\begin{align*}
|\nu_1|(x) 
& \leq (2\pi^2)^{-1} g^4 \int_{r\geq\mu} |x - (\cdot)|^{-3} 
\psi^2 \nu dvol_5\tag{131} \\
& \leq (2\pi^2)^{-1} \mu^2 e^{-2g^{1/4}/\mu}  
z(\nu) \int_{r\geq\mu} |x - (\cdot)|^{-3} r^{-5+\ve}dvol_5.
\end{align*}
Here, $dvol_5$ denotes the Euclidean volume element on $\bR^5$.  
Now, the right most integral in (131) is no smaller than $(2\pi^2)\ve^{-1}
|x|^{-3+\ve}$, and so 
$$
|x|^{3-\ve}|\nu_1|(x)\leq\ve^{-1}\mu^2 e^{-2g^{1/4}/\mu} z(\nu) .\eqno(132)
$$
Then, as $z(\nu) \leq z(\nu_0) + z(\nu_1)$, this last inequality implies that
$$
\sup_{r\geq\mu} r^{3-\ve} |\nu_1| =  z(\nu_1) 
\leq  2 \ve^{-1} \mu^2 e^{-2g^{1/4}/\mu}  z(\nu_0) \eqno(133)
$$
as long as $\ve\geq 2 \mu^2e^{-2g^{1/4}/\mu} $.  	
	
This last point with (129) implies that there is a $g$-independent
constant $c \geq 1$ such that  when $g \geq c$ then
$$
\al |_{r=2\mu} \geq c^{-1} g^{1/2} \sin \ta.\eqno(134)
$$
This understood, (9) plus the maximum principle implies that 
$$
\al\geq c^{-1} g^{1/2} r^{-1} \sin \ta\eqno(135)
$$
at all $r \geq 2\mu$.

\section*{j)  The uniqueness of solutions}

The purpose of this section is to consider the following
question:  Are there two distinct,  
non-negative functions which are both absolute minima of $\ce_0$ in (10)?
Here is the answer: 

\begin{proposition}			
For any $g > 0,$ the functional $\ce_0$ has a unique, non-negative
minimizer. 
\end{proposition}

The remainder of this section is occupied with the proof of this proposition.

\medskip\noindent
{\it Proof of Proposition} 13. \ 
The first step in the proof is to elaborate on the conclusions of Lemma 
7.  This is provided by	

\begin{lemma}		
If an absolute minimizer, $\al$, to $\ce_0$ is somewhere positive, 
then $\al$ is bounded from  below where $r \geq 1$ by a constant, 
non-zero multiple of $r^{-2} (r - 1) \sin \ta$.
\end{lemma}

\noindent
{\it Proof of Lemma} 14. \   
First, let $\nu\equiv(r^2 \sin \ta)^{-1} \al$ 
and consider $\nu$ and $\psi$ as in Step 7 of the previous 
section to be functions on $\bR^5$ via identifications 
$r = (x_1^2 + \cdots + x_5^2)^{1/2}$ and 
$\ta ={\rm Arccos}\,(x_5/r)$ on $\bR^5$.  
As previously noted, the function $\nu$ obeys
\begin{itemize}
\item
$\nu = 0$    where   $r \leq 1$. \hfill (136)
\item
$\De_5 \nu- g^4 \psi^2 \nu= 0$    where   $r > 1$.
\item
$\nu$ is square integrable over $\bR^5$.
\end{itemize}
As before $\De_5$ denotes the standard Laplacian on $\bR^5$.  
Now, when $y \in \bR^5$
has $|y| > 1$, use $G(\cdot\, ;\, y)$  
denote the Green's function on the complement of the unit ball in
$\bR^5$ with Dirichelet boundary  
conditions on the surface of the ball and with pole at $y$.  Thus, 
$$
G(x; y) = (2\pi^2)^{-1}(|x - y|^{-3}-||y|x - y/|y||^{-3}).\eqno(137)
$$
By inspection, $G(x; y) > 0$ where $|x| > 1$, and $\pa_rG(\cdot\,;\,
y) |_x > 0$ where $|x| = 1$.  This said, then $\nu$ in (136) can be
written in terms of $G$ as
$$
\nu(x) = g^4 \int G(x; y) (\psi^2 \nu)|_y d^5y .\eqno(138)
$$
The claims in the lemma follows directly from this representation of $\nu$.

With Lemma 14 in hand, suppose now that $\al$ is an absolute minimizer of
e0 and is not identically zero.  Let $\al'$ be any other function in
the domain of $\ce_0$.  By virtue of Lemma 14, the function $\al'$ can be
written as $\al' = h \al$, where $h$ is smooth where $r > 1$ and bounded on
sets where $r$ is bounded.  As is demonstrated below, $\ce_0(\al')$ can be
written in terms of $h$ as
\begin{align*}
\ce_0(\al') &= \ce_0(\al) + 2^{-1} g^{-2} 
\int(h_r^2 + r^{-2} h_{\ta}^2) r^{-2} \al^2 d^3y\tag{139}\\ 
&\quad+ 2^{-1} g^2 \int(h^2 - 1) r^{-2} \al^2 \psi (\psi - \psi') d^3y ,
\end{align*}
where $\psi'$ is the solution to the $\al'$ version of (8).  
To see the significance of (139), first note that the 
primed and unprimed versions of (8) together imply that
$$
-\De(\psi - \psi) + r^{-2} \al^{\prime 2} (\psi-\psi') 
+ r^{-2} (1 - h^2) \al^2 \psi = 0 .\eqno(140)
$$
Now, multiply both sides of this last equation by $(\psi-\psi')$ and then
integrate over $\bR^3$.  An integration by parts results in the equality
\begin{multline*}
\int(|\na(\psi-\psi')|^2 + r^{-2} \al^2 (\psi-\psi')^2 d^3y \\ 
+ \int r^{_2} (1 - h^2) \al^2 \psi (\psi-\psi') d^3y  = 0 .\tag{141}
\end{multline*}
This last equation precludes a negative value for the second term on
the right hand side of (141); in fact, said term is positive unless $\psi 
=\psi'$ and thus $h = 1$.  Hence, $\ce_0(\al') > \ce_0(\al)$ 
unless $\al' =\al $.
	
By way of tying loose ends, what follows next is the derivation of
(139).  To begin, multiply both sides of the equation in the middle
point of (9) by $h^2 \al$ and then move one factor of $h$ through the various
derivatives on $\al$ to write the leading order derivatives in terms of
$\al' = h\al$.  This done, integrate the result over $\bR^3$ and then
integrate once by parts to see that the difference between the $\al'$
and $\al$ versions of the first integral in (10) is equal to
$$
2^{-1} g^{-2} \int(h_r^2 + r^{-2} h_{\ta}^2) r^{-2} \al^2 
+  2^{-1} g^2 \int(h^2 - 1) r^{_2} \al^2 \psi^2 .
\eqno(142)
$$
Next, note that the difference between $\al'$ and $\al$
versions of the second integral in (10) is equal to
$$
\int (\psi' - \psi) \rho .\eqno(143)
$$
The latter is equal to the integral that is obtained 
by multiplying both sides of the equation in the top 
point of (8) by $(\psi' - \psi)$ and then integrating over $\bR^3$.  
Two applications of integration by parts then 
equates (143) with
$$
\int (\psi (-\De(\psi'- \psi) + r^{-2} \al^2 (\psi-\psi'))) d^3y.
\eqno(144)
$$
To utilize this last identity, note that (140) can be rewritten as
$$
-\De(\psi-\psi') + r^{-2} \al^2 (\psi-\psi') 
+ r^{-2} (1 - h^2) \al^2 \psi' = 0 .\eqno(145)
$$
This last equation implies that the integral in (143) is equal to
$$
\int r^{-2} (1 - h^2) \al^2 \psi\psi' d^3y .\eqno(146)
$$
Thus, the contribution to $\ce_0(\al') - \ce_0(\al)$ from the second
integral on the right side of (10) is equal to the expression in (146)
times $2^{-1} g^2$.  Adding the latter to (142) produces the desired right
hand side of (139).

\section*{k)  The proofs of Theorems 1 and 2}

The purpose of this final section is to present the proofs of
the the two theorems in the introduction.  As the proof of Theorem 2
requires little more than the collation of results from the preceding
sections, it is given first.

\medskip\noindent 
{\it Proof of Theorem} 2. \ 
The fact that minimizer, $\al$, of $\ce_0$ is unique up 
to multiplication by $\pm 1$ is proved as
Proposition 13.  Lemma 3 proves that the minimizer is $\al\equiv 0$ when $g
\leq (6\pi)^{1/2}$ and not so otherwise.  The bounds in (6) for $\ce_0(\al)$
follow from Propositions 4 and 8 using (14).  The bounds in (7) also
follow from these propositions with the help of (14) and (17).  The
positivity of $\psi$ asserted by the first point of Theorem 2 follows by
applying the maximum principle to the first equation in (8).  Lemma 14
asserts the conclusion of the second point of Theorem 2, and the third
point of Theorem 2 is a restatement of the conclusions of Proposition
9.  The fact that $r\psi$ has a limit as $r \ra\iy$ 
equal to $\surd 2 g^{-2} e_0$ with $e_0 \leq 1$ is proved 
above as Proposition 6.  The fact that $e_0 > 0$ is proved in Section g.  
The bound $e_0 e^{-g^{1/4}/c}$ follows from the assertion
in Part a of the third point of Theorem 2 using (8) and the maximum
principle.  Meanwhile, the assertion from Part b that 
$\al\sim c_0 r^{-[(9-8e_0^2)^{1/2}-1]/2} \sin \ta$ with $c_0$ bounded 
is discussed in Section g.  

To complete the proof of the final point of Theorem 2, 
here is the argument for the assertion
that $r |m_{\psi}|$ is bounded: First, write $\psi$ using (8) as
$$
\psi(x) = (4\pi)^{-1} |x|^{-1} 
- (4\pi)^{-1} \int |x - (\cdot)|^{-1} \al^2 r^{-2} \psi.\eqno(147)
$$
As $\psi\leq r^{-1}$ and $|\al| \leq r^{-3/4}$ 
at large $r$, it follows that both $\al^2r^{-2}\psi$
and $\al^2 r^{-1}\psi$ are integrable.  This  said, the bound on 
$|m_{\psi}|$ follows from (147) since $|x - y|^{-1} - |x|^{-1}
= \co(|y|/|x|^2)$ when $|y| < |x|/8$.    
	
Finally, here is the argument for the bounded behavior of
$r|m_{\al}|$: Write $\al$ in terms of the function $\nu$ that appears
in (138).  It follows from (138) and the bound just proved for
$m_{\psi}$ that the function $m_{\al}$ obeys a fixed point equation of
the form $m = T(m)$ with
$$
T(m) =e^{-g^{1/4}/c} |x|^2 \int_{|y|>|x|/4} G(x, y) 
|y|^{-4-[(9-8e_0^2)^{1/2}-1]/2} m(y) d^5y + h(x).  \eqno(148)
$$ 

Here, $h$
is a function that satisfies $|h(x)| \leq c |x|^{-1}.$  Now, $T(m)$ is
uniformly contracting on the Banach space of bounded functions which
vanish where $|x| \leq 1$, and so it has a unique fixed point on this
space.  This fixed point is $m_{\al}$.  This said, introduce the Banach space
b consisting of those functions m that vanish where $|x| \leq 1$ and are
such that $|x| |m|$ is bounded.  Here, the norm of $m \in \cb$ is the supremum
on $\bR^3$ of $|x| |m|$.  As $T$ also maps $\cb$ to itself as a uniformly
contracting operator, so it has a unique fixed point in $\cb$.  Since the
functions in $\cb$ are bounded on $\bR^3$, the latter fixed point is the same
as the former; thus $m_{\al} \in\cb$.

\medskip\noindent {\it Proof of Theorem} 1. \ The assertion that the
coulomb solution is the minimizer of the small $g$ versions of $\ce$
is proved last.  The coulomb solution is not the absolute minimum of
any $g > (6\pi)^{1/2}$ version of $\ce$ since it is not the minimizer
of the same $g$ version of $\ce_0$.  As for the other assertions, the
upper bound in (6) follows from the analogous, $\ce_0$ version.
Meanwhile, the lower bound in (6) is a consequence of the first point
and the lower bound in either the second or third points of (7).  In
this regard, the argument for the first point of (7) appears just
after Theorem 2.  Meanwhile, the upper bounds in the second and third
points of (7) follow given the first point and the upper bound in (6).
As the lower bounds in the second and third points of (7) are proved
momentarily, consider in the mean time the arguments for the fourth
point.  In particular, the pointwise bound on $|\Psi_A|$ by
$|\Psi_{A=0}|$ follows via the maximum principle since $u \equiv
|\Psi_A|$ satisfies the differential inequality $-\De u \leq \rho_0$
where $\rho_0$ is zero where $r \geq 1$ and equals $3/4\pi$ where $r
\leq 1$.  The bound on $|\Psi_A|$ by $c g^{-1/2} r^{-1}$ where $r \geq
2$ is proved by essentially the same argument that proves the final
point of Proposition 4.

Given the upper bounds in the second and third points of Theorem 2,
the corresponding lower bounds in these points follow as corollaries
to

\begin{proposition}			
There is a constant $c \geq 1$ with the following significance: 
Suppose that $A \in \ca,$ that $\ga\geq 1$ and that the corresponding 
$\Psi_A$ obeys $\int_{r\geq 1} |\na_A\Psi_A|^2 \leq\ga^{-1}.$  
Then, $\int_{1\leq r\leq c/\ga}|B_A|^2 \geq c^{-1} \ga^3.$
\end{proposition}

\noindent
{\it Proof of Proposition} 15. \   
Before starting, note that the argument is very much the same as the
one above that proved Proposition 8.  In any event, to start, let $\psi$
now denote $-2$ trace$\,(\tau^1\Psi_A)$.  
The argument given in Steps 1-2 of the proof of Proposition 8 
find $\ga$-independent constants $c_1$, $c_2$ and $K$, all
greater than 1, and a unique $\ka\geq 1/K$ such that the following is
true: Let c denote the space of Sobolev class L21 functions on $\bR^3$ that
vanish where $r \geq 1 + c_2 \ga^{-1}$.  Then,
\begin{align*}
&3 (8\pi)^{-1} \int_{r\leq 1}\psi\tag{149}\\
&= \sup_{u\in\cC} \left\{ 3 (4\pi)^{-1} \int_{r\leq 1} u  
-2^{-1}\left(\int|\na u|^2+\ka^2\ga^2\int_{r\geq 1}r^{-2}
u^2\right)\right\}.
\end{align*}
holds when $\ga\geq 64 \pi c_1$.  Moreover, the supremum on the right
hand side of (149) is achieved by the  function $u_{\ka}$ from (84) 
where $p, p'$ and $d$ are by the formulas in (83)
after replacing $g$ with $\ga$.   This understood, note that 
\begin{align*}
& 3 (8\pi)^{-1} \int_{r\leq 1} \psi \tag{150}\\
& = \sup_u \left\{3 (4\pi)^{-1} \int_{r\leq 1}(-2\,{\rm trace}\,
(\hat{u}\tau^{-1}))-2^{-1} \int |\na_A\hat{u}|^2 \right\},
\end{align*}
where the supremum is taken over all $su(2)$-valued 
functions $\hat{u}$ with both $\na_A\hat{u}$ and $r^{-1} \hat{u}$ square 
integrable.  This understood, remark next that 
$\hat{u}\equiv u_{\ka} \tau^1$ is such a function and as 
$$
\na_A\hat{u} = \na u_{\ka} \tau^1 + [A, \tau^1] u_{\ka},\eqno(151)
$$
so (150) implies that
$$
\int |[A, \tau^1]|^2 u_{\ka}^2 \geq \ka^2 g^2 
\int_{r\geq 1} r^{-2} u_{\ka}^2 .\eqno(152)
$$
	
As in the derivation of (93), this last inequality implies the
existence of a $\ga$-independent constant $C \geq 1$ such that
$$
\int_U |A|^2 \geq C^{-2} \ga ,\eqno(153)
$$
where $U$ is the portion of $\bR^3$ where $1\leq r\leq 1+c_2\ga^{-1}$.
Since $A|_{r=1} = 0$, this last equation implies that $\int_U
|\pa_rA|^2 \geq C' \ga^3$ where $C'$ is a positive, $\ga$-independent
constant.  To finish the argument, note that every element in $\ca$ is
gauge equivalent to some 1-form that annihilates the vector field
$\pa_r$; and for such $A$, the inequality $|B_A| \geq |\pa_rA|$ holds.

The proof of Theorem 1 now lacks only the justification of the
assertion that the Coulomb solution is the minimizer when $g$ is less
than the given bound.  To argue this point, agree, first of all, to
consider only those $su(2)$ valued 1-forms $A \in \ca$ that annihilate the
vector field $\pa_r$.  As just noted at the end of the proof of the
preceding proposition, every element in $\ca$ is gauge equivalent to such
a 1-form.  (In fact, the gauge transformation is unique if required to
be the identity on the radius 1 ball.)  

Now, to start the argument, use (1) to conclude that
$$
\int |\na_A\Psi_A|^2 = 3 (4\pi)^{-1} \int_{r\leq 1} -2\, 
{\rm trace}\,(\tau^1\Psi_A).\eqno(154)
$$
Meanwhile, (1) also implies that
$$
\int\lef \na_A\Psi_{A=0}, \na_A\Psi_A\ri 
=3(4\pi)^{-1}\int_{r\leq 1}-2\,{\rm trace}\,(\tau^1\Psi_{A=0})=2\ce(0).
\eqno(155)
$$
Here, $\lef , \ri$ is shorthand for the inner product 
on $su(2)$-valued 1-forms. Now, as $A$ has no $dr$  
component and $\na\Psi_{A=0}$ has only a dr component, 
the left hand side of (155) is equal to
$$
\int\lef-\De_{A=0}\Psi_{A=0}, \Psi_A\ri 
+ \int\lef[A, \Psi_{A=0}], \na_A\Psi_A\ri .\eqno(156)
$$
Furthermore, as $-\De_{A=0}\Psi_{A=0} = \rho$, 
this last equation and (154) imply that
$$
2^{-1} \int |\na_A\Psi_A|^2 - \ce(0) 
= - 2^{-1} \int\lef[A, \Psi_{A=0}], \na_A\Psi_A\ri .\eqno(157)
$$
To proceed, use the lack of $dr$ component in $A$ 
and the lack of other components in $\na\Psi_{A=0}$ 
to equate the right hand side of (157) with
$$
- 2^{-1} \int\lef[A, \Psi_{A=0}], \na_A(\Psi_A- \Psi_{A=0})\ri 
- 2^{-1} \int|[A, \Psi_{A=0}]|^2 .\eqno(158)
$$
This done, the triangle inequality finds
\begin{align*}
&2^{-1} \int |\na_A\Psi_A|^2 - \ce(0)\tag{159}\\ 
&\geq - 4^{-1} \int|\na_A(\Psi_A - \Psi_{A=0})|^2 
- (3/4) \int|[A, \Psi_{A=0}]|^2 .
\end{align*}
The first term on the right hand sider of (159) is awkward for the
purposes at hand.  To  replace it, 
note that (1) and it's $A = 0$ version imply that
$$
-\De_A(\Psi_A - \Psi_{A=0}) = \na_A\bullet[A, \Psi_{A=0}]\eqno(160)
$$
where $\bullet$ signifies the contraction of 1-form indices.  
Here again, use has been made of the lack of a $dr$ component of $A$ 
and the lack of the other components of $\na\Psi_{A=0}$.  
Contract both sides of this last equation with 
$\Psi_A - \Psi_{A=0}$, integrate the result over $\bR^3$, 
integrate by parts and, finally, employ the 
triangle inequality to discover that
$$
\int|\na_A(\Psi_A - \Psi_{A=0})|^2 \leq \int|[A, \Psi_{A=0}]|^2 .
\eqno(161)
$$
Thus, (159) implies that
$$
2^{-1} \int|\na_A\Psi_A|^2 - \ce(0) 
\geq - \int|[A, \Psi_0]|2 \geq - (4\pi)^{-2} \int r^{-2} |A|^2 .
\eqno(162)
$$
	
Meanwhile, as the lack of $dr$ component in $A$ also implies 
that $|B_A| \geq |\pa_rA|$,
so it follows, after appeal to the first line in (21), that
$$
\ce(A) - \ce(0) \geq 8^{-1} g^{-2} \int r^{-2} |A|^2 
- (4\pi)^{-2} g^2 \int r^{-2} |A|^2 .\eqno(163)
$$
Thus, $\ce(A) \geq \ce(0)$ when $g^2 \leq \surd 2 \pi$.

\end{document}